\documentclass[12pt]{iopart}
\usepackage[utf8]{inputenc}
\pdfoutput=1

\usepackage{graphicx}
\usepackage{amssymb}
\usepackage{subfig}
\usepackage{url}
\usepackage{cite}

\newcommand{\D}{\mathrm{d}}
\newcommand{\Li}{\mathrm{Li}}

\begin{document}

\title[Order in a moving condensate]{Maintenance of order in a moving strong condensate}
\author{Justin Whitehouse, Andr\'{e} Costa, Richard A Blythe, Martin R Evans}

\address{SUPA, School of Physics and Astronomy, University of Edinburgh, Mayfield Road, Edinburgh EH9 3JZ, UK}

\begin{abstract}
We investigate the conditions under which a moving condensate may exist in a driven mass transport system. Our paradigm is a minimal mass transport model  in which $n-1$ particles move simultaneously from a site containing $n>1$ particles to the neighbouring site in a preferred direction. In the spirit of a Zero-Range process the rate $u(n)$ of this move  depends only on the occupation of the departure site. We study a hopping rate $u(n) = 1 + b/n^\alpha$ numerically and find a moving strong condensate phase for $b > b_c(\alpha)$ for all $\alpha >0$. This phase is characterised by a condensate that moves through the system and comprises a fraction of the system's mass that tends to unity. The mass lost by the condensate as it moves is constantly replenished from the trailing tail of low occupancy sites that collectively comprise a vanishing fraction of the mass. We formulate an approximate  analytical treatment of the model that allows a reasonable estimate of $b_c(\alpha)$  to be obtained. We show numerically (for $\alpha=1$) that the transition is of mixed order, exhibiting exhibiting a discontinuity in the order parameter as well as a diverging length scale as $b\searrow b_c$.
\end{abstract}


\maketitle 

\section{Introduction}

In nonequilibrium statistical physics, condensation is used as a general term to describe the localisation of a finite fraction of some quantity---typically mass---in a wide variety of fundamental models of dynamical processes. These include the flow of wealth \cite{Burda2002}, traffic flow \cite{OLoan1998, Chowdhury2000, Levine2004a, Kaupuzs2005}, and the formation of hubs in complex networks \cite{Krapivsky2000, Angel2005}. The archetypal model of this class is the Zero-Range Process (ZRP) \cite{Evans2000, Evans2005}. In this minimal model, single units of mass hop between sites at a rate which is a function only of the total mass on the site they are leaving (hence the name `zero-range'). Furthermore this model satisfies the conditions required for the steady state to factorise, which simplifies the analysis of its condensate phase \cite{Evans2004, Majumdar2005a, Evans2006}. Given an appropriate choice of the hopping rate $u(n)$, which decreases suitably slowly with $n$, this process alone is enough to create a \emph{static} condensate phase in which a finite fraction of the total mass of the system occupies a single site. For the case $u(n)= 1 + b/n^\alpha$ the transition has been extensively studied. When $\alpha =1$  there is a  critical value  of the parameter $b$,  $b_c=2$, above which a condensation transition occurs when particle density $\rho$ exceeds a critical value $\rho_c=1/(b-2)$. For $\rho <\rho_c$, the system is in a fluid phase where the mass is  evenly distributed across sites, whereas for $\rho > \rho_c$ a condensate emerges. For  $\alpha < 1$ a condensation transition occurs for all $b$ \cite{Evans2000}.

There are certain cases in which the nature of the condensate phase is different from the `standard' condensation described above. For example: the fraction of the total system mass in the condensate  can be equal to 1, creating a strong condensate \cite{Jeon2000,Jeon2010}; the condensed phase can exhibit a subextensive number of smaller mesocondensates \cite{Schwarzkopf2008} or an extensive number of finite-sized quasi-condensates \cite{Thompson2010}. Also, it should be noted that the existence of a condensate phase is not unique to models based on the ZRP or with factorised steady states. For example, a non-Markovian simple exclusion process has been shown to exhibit an immobile condensate phase \cite{Concannon2014}.

In all these examples, the condensates are static: they reside at the same point in space for a long period until dissolving through a large fluctuation and reforming elsewhere \cite{Godreche2003}. However, in physical settings moving condensates or aggregates are often observed, for example in traffic jams \cite{Lighthill1955}, gravitational clustering \cite{Silk1978}, sedimentation \cite{Horvai2008} and droplet formation \cite{Family1989}. In general, moving condensates are less well understood than the static variety and it is unclear what the physical mechanisms are that will allow the maintenance of the condensate.

In this work we investigate conditions under which a condensate may maintain its order as it moves through the system. To understand why this is a pertinent question we first review how condensates move in a variety of simple model systems related to the ZRP.

In an important early contribution, Majumdar, Krishnamurthy and Barma \cite{Majumdar1998} introduced a chipping model in which all the mass from a site can move, or `diffuse', to an adjacent site. Additionally, a single unit of mass can `chip' off from the departure site and hop to an adjacent site.  For symmetric diffusion of the mass, a condensed phase was observed.  However for asymmetric diffusion, which leads to a condensate moving on average, a careful analysis revealed that, although a condensate is still observed on a finite system \cite{Majumdar2000}, the critical density at which the condensation transition occurs diverges in the thermodynamic limit \cite{Rajesh2001}. This is because the chipping process (in one-dimension) dissipates clusters faster than the diffusion process creates them \cite{Rajesh2002}. However subsequent work on a chipping model with a chipping rate of the classical zero-range type, $u(n) = 1 + b/n$, suggests that a condensate may be possible for large enough $b$ with the critical value  $b_c$ somewhere close to two \cite{Levine2004}.

Meanwhile, Hirschberg \textit{et al.} have investigated what kinds of dynamical processes will permit a moving condensate phase using variants of the ZRP with non-Markovian hopping rates \cite{Hirschberg2009,Hirschberg2012} and with hopping rates affected by spatial correlations \cite{Hirschberg2013}. Both models exhibit a condensate phase which drifts with a finite, nonvanishing velocity. In the former, temporal correlations between departure and arrival sites allow the formation of a condensate over two adjacent sites, which then moves with a `slinky'-like motion through the system. In the latter, the effect of spatial correlations is that the condensate also moves with a slinky-like motion, but with certain differences in the details depending on the values of certain hopping parameters. Condensate motion of a similar slinky nature  has also  been observed in a totally asymmetric model  \cite{Waclaw2012, Evans2014} in which the hopping rate is a monotonically increasing function of the mass at both departure and arrival sites. The condensation found in this model is found to be ``explosive'' in as much as the condensate moves with a superextensive velocity and forms instantaneously in an infinite system.

Taken together these studies pose the intriguing question that we pursue here: What are the key dynamical processes that  permit a  moving condensate phase, and which processes will destabilise the phase?  This question is of broader relevance to phenomena such as flock formation and schooling of fish \cite{Toner2005}.

In this work, we  introduce a minimal mass-transport model in the spirit of the ZRP which allows large-mass hopping events.
The new feature is the incorporation of a `backchip' move described below and illustrated in \fref{fig:stack_hop}. We find that the model exhibits a moving condensate phase with a distinctive mechanism of formation and maintenance. Most notably, the moving condensate is a \emph{strong} condensate in the sense that a fraction tending to one of the particles occupy a single site. This condensate travels through the system followed by a tail of low occupancy sites that collectively comprise a vanishing fraction of the mass. As we show below, the dynamics of the mass within this tail is responsible for maintaining the structure of the condensate.  Using numerical simulations, we find that above a critical value for a rate parameter $b$, and at all densities, a strong condensate forms. Numerically we are able to classify the transition as being of mixed order, exhibiting features of both first and second order phase transitions. We further provide an approximate theory of the mechanism which gives a reasonable prediction for the critical value, and also discuss the behaviour of the system below this critical rate.

\section{Model and motivation}

To motivate the specific features of our model let us first consider the limit of zero chipping rate in the models of \cite{Majumdar2000,Levine2004}. In the absence of any chipping, the dynamics is simply diffusion combined with irreversible aggregation. The stationary state of this process on a finite system comprises a single condensate containing all the system's mass. The work of \cite{Rajesh2001,Rajesh2002} has shown that the condensate is unstable to the effect  of single particles chipping  away from an aggregate with rate $u=1+b/n$,  where $n$ is the number of particles contained within the aggregate, unless $b$ is greater than a critical value $b_c >0$. Therefore it is of interest to consider what other perturbations may destabilise the condensate exhibited in diffusion with irreversible aggregation.

\begin{figure}
  \centering
  \subfloat[]{\label{fig:single_hop} \includegraphics[width=0.25\linewidth]{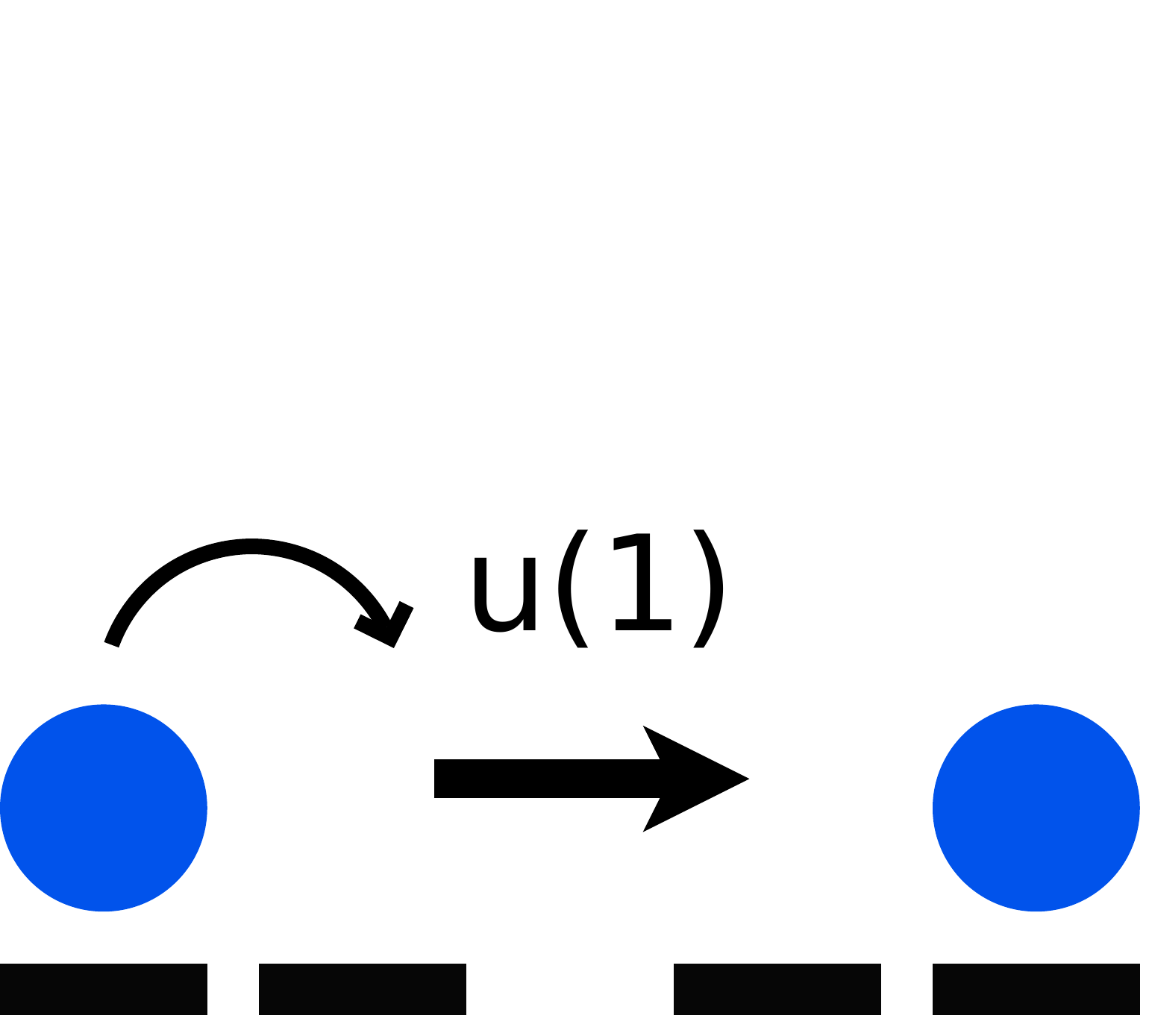} }\hspace{0.1\linewidth}
  \subfloat[]{\label{fig:stack_hop} \includegraphics[width=0.25\linewidth]{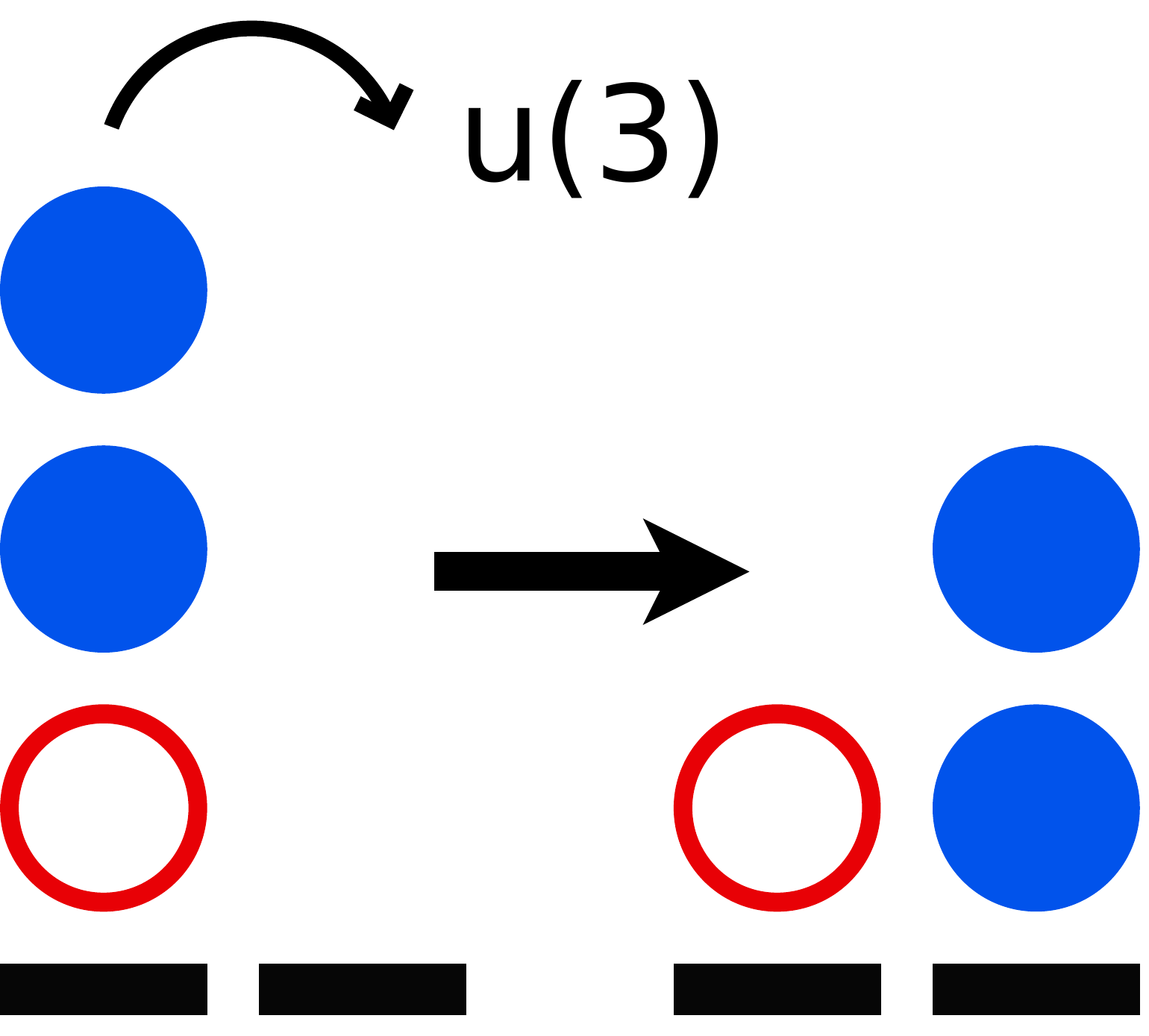} }
  \caption{The elementary dynamical processes. (a) Hop: When a site contains only one particle, this particle hops onto the next site with rate $u(1)$ and leaves behind an empty site. (b) Backchip: When a site contains $n$ (here, 3)  particles, $n-1$ of these particles move together with rate $u(n)$ onto the next site and leave behind a \emph{single} particle.}
  \label{fig:hops}
\end{figure}

Here we consider a different perturbation of the diffusion with aggregation dynamics: when the aggregate moves forward it leaves one unit of mass behind---see \fref{fig:hops}. We retain the zero-range feature that the rate of movement $u(n)$  only depends  on the number of particles within the aggregate. 

To be precise, we consider a system of $N$ particles upon a one-dimensional lattice of $L$ discrete sites. We are interested in the large $N$,$L$ behaviour where the density $\rho = N/L$ is fixed. In the case where the site has occupancy $n > 1$, a `backchip' takes place: $n-1$ of the particles move to the next site, and leave behind a single particle (\fref{fig:stack_hop}). 
This occurs
with rate with rate $u(n)$ given by
\begin{equation}\label{eq:hop_rate}
  u(n) = 1 + \frac{b}{n^\alpha} \;,
\end{equation}
where $n$ is the total number of particles on the departure site. With this form, larger values of the rate parameter $b$ bias the dynamics towards faster hopping from less occupied sites, causing larger groups of mass to move slowly in comparison. 

The dynamical rule has of course to be modified for a single unit of mass at a site, i.e. when the site's  occupancy is $n=1$. This single particle moves to the next site and leaves behind an empty one (\fref{fig:single_hop}). We take the rate of this `hopping' process to be $u(1)$.

Our choice of hop rates allows us to compare our model to the standard formulation of the ZRP where only a single particle can hop at a time.  This type of hop is often referred to as a `chip', as in \cite{Majumdar1998, Majumdar2000, Rajesh2001, Rajesh2002}, and is in some sense symmetric to our definition of a backchip.  The name `chip' can be conceptually understood in the context of a single mass unit chipping off from a site with large occupancy. We first study the case where $\alpha = 1$, as it can be shown exactly that with the hop rate given in \eref{eq:hop_rate} and $\alpha=1$ the standard ZRP undergoes a condensation transition at $\rho_c=1/(b-2)$ for $b >2$ \cite{Evans2005}.

\section{Monte Carlo simulations}

\begin{figure}
  \centering
  \subfloat[]{\label{fig:strongcon} \includegraphics[width=0.35\linewidth]{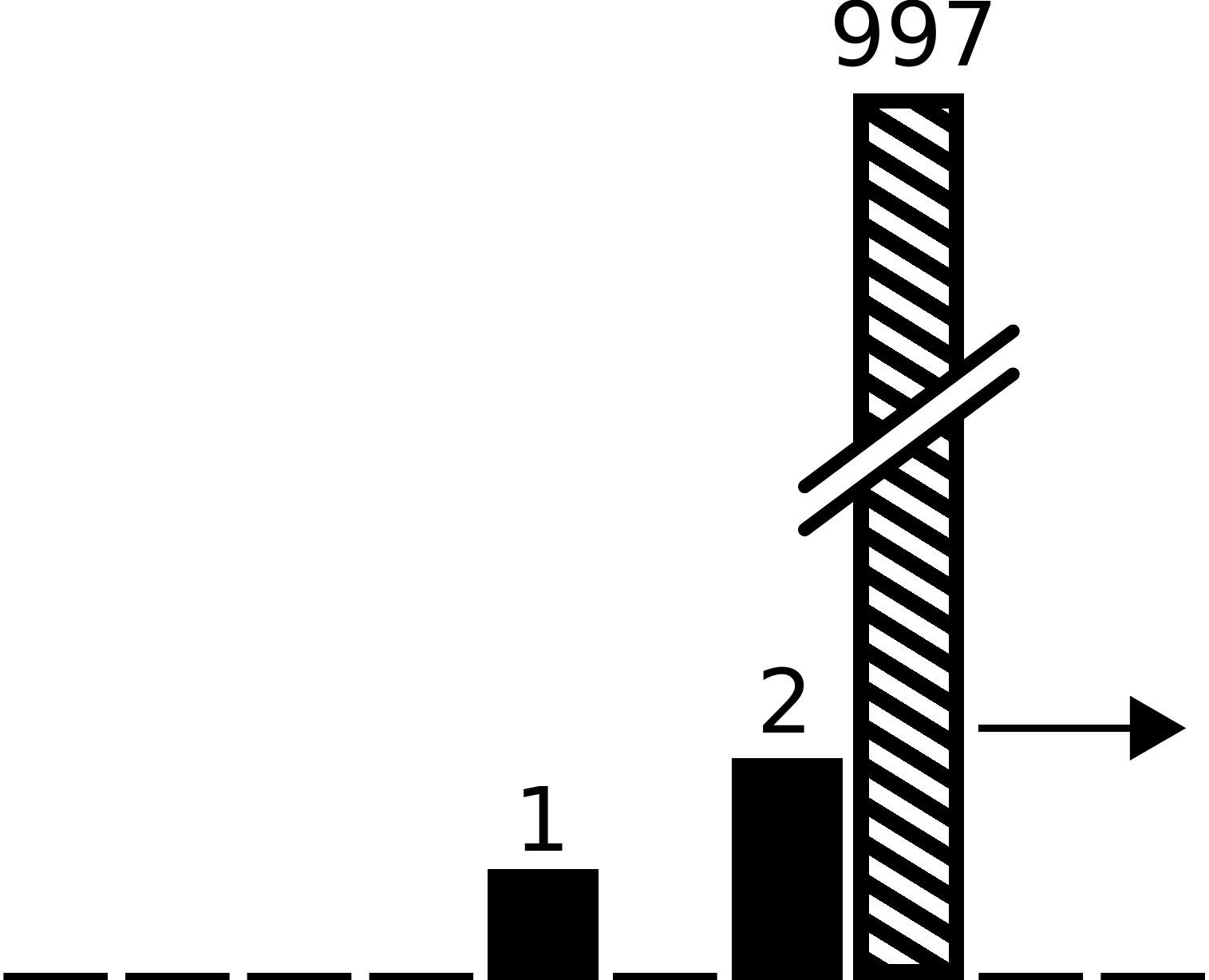} }\hspace{0.1\linewidth}
  \subfloat[]{\label{fig:strongconprof} \includegraphics[width=0.35\linewidth]{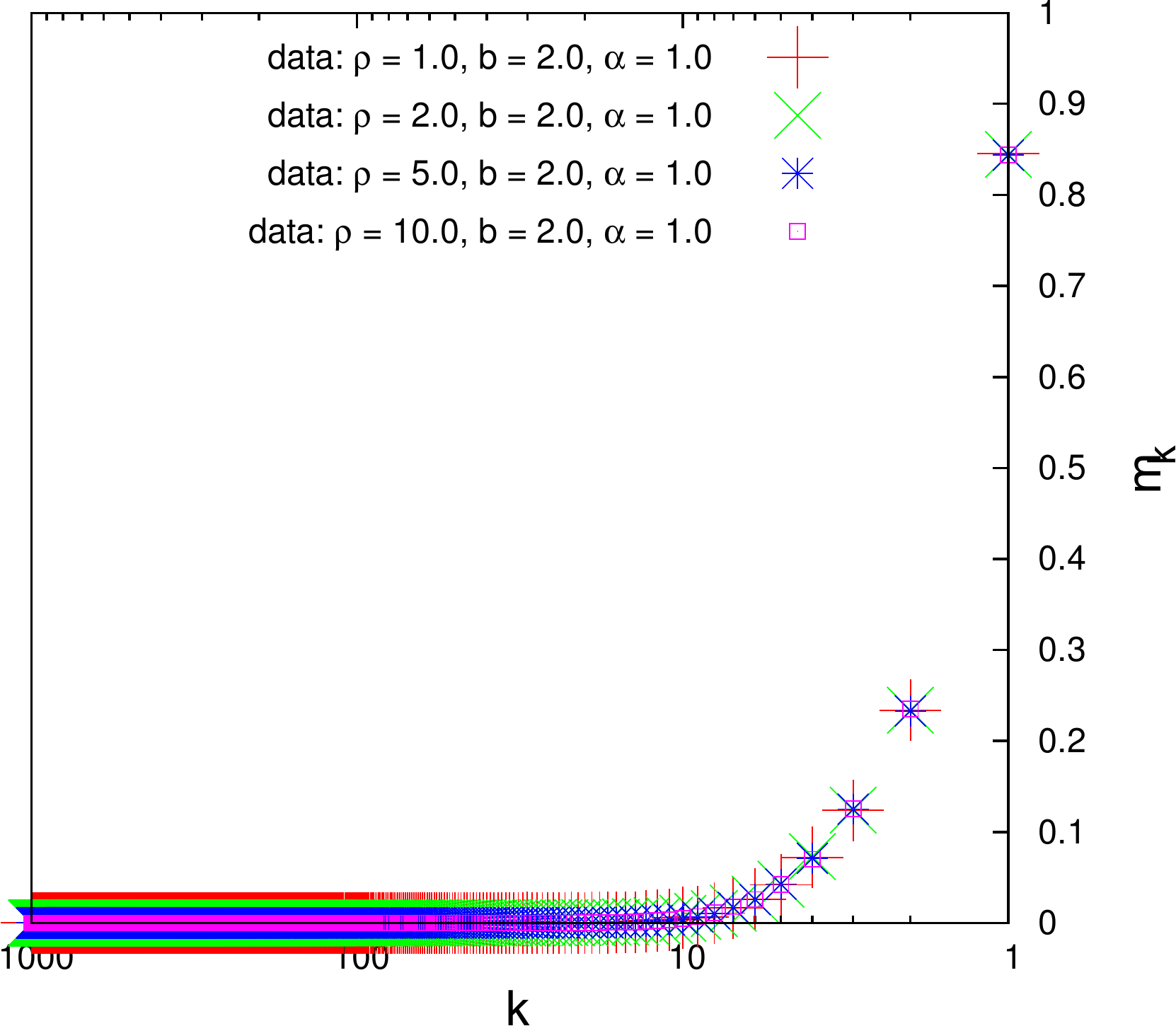} }
  \caption{Illustration of the condensate phase. (a) A typical configuration of the system in the strong condensate regime, sketched using data taken directly from a simulation with $L = 1000$, $N = 1000$, $b = 2.0$ and $\alpha = 1.0$. The columns represent the mass occupying the site, with the exact size shown above, and the direction of motion is indicated by the arrow.
  (b) The average mass $\overline{m_k}$ at a site $k$ sites \emph{behind} the condensate for $\rho = 1,2,5,10$. In the strong condensate regime all of the system's mass is typically in the condensate, with any remaining units of mass trailing closely behind. }
  \label{fig:config}
\end{figure}

\begin{figure}
  \centering
  \includegraphics[width=0.6\linewidth]{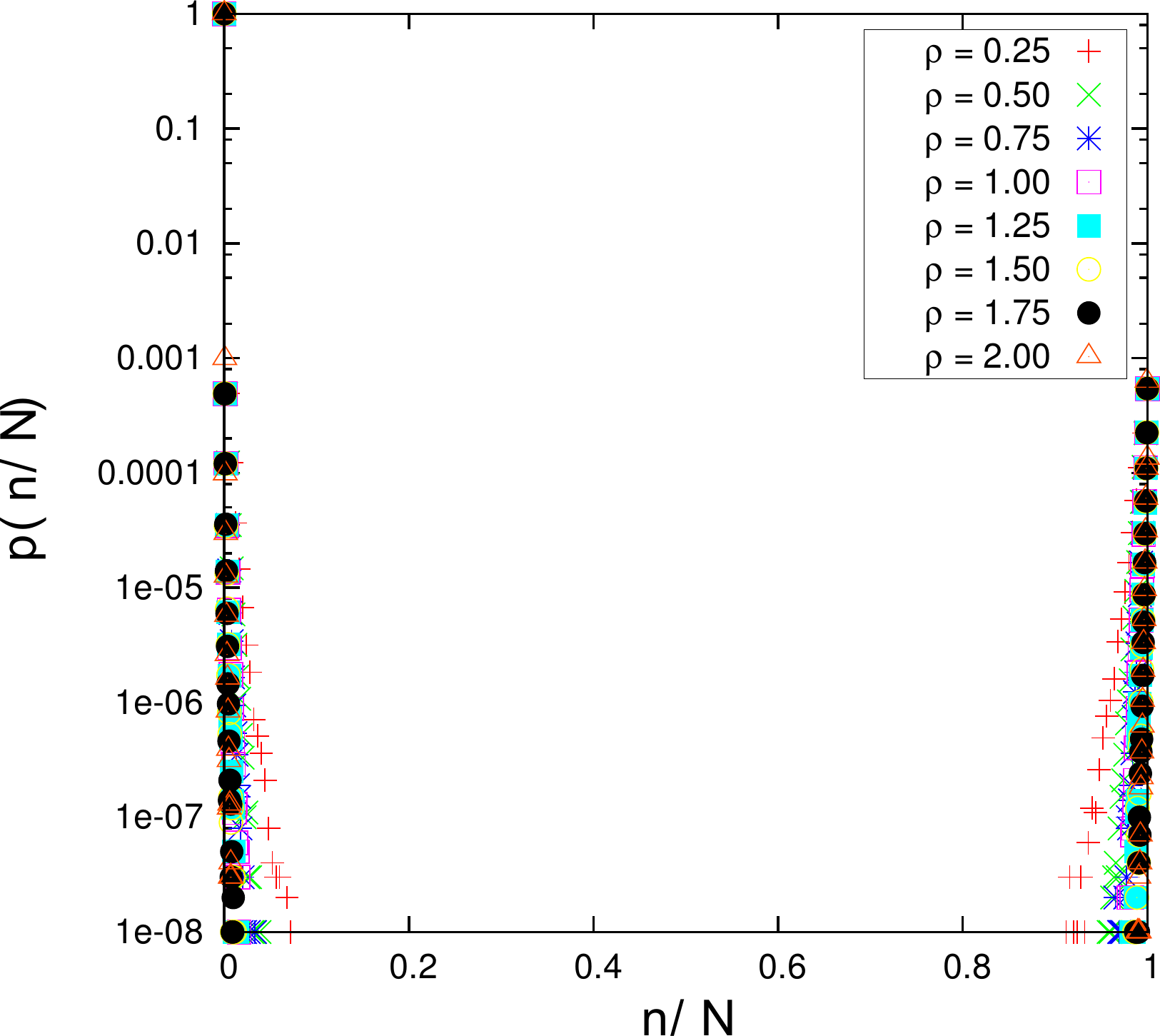}
  \caption{Plot of the distribution $p(n)$ of site occupancies $n$ in the strong condensate regime at $b = 2.0$ for a system of size $L = 1000$. $N$ is the total number of particles in the system. The points at and near $n/N=1$ indicate the presence of s strong condensate which contains effectively all of the system's mass.}
  \label{fig:dist_plot_vary_rho}
\end{figure}

We implemented Monte Carlo simulations of the system on a one-dimensional periodic lattice. From these we see that, above a critical value of $b$, the system exhibits a \emph{strong} condensate, where almost all the mass occupies a single site, which moves through the system. This is immediately followed by a short tail of sites with very low occupancy, leaving all other sites empty (\fref{fig:config}). This behaviour can also be seen from the plots of the site occupancy distribution (\fref{fig:dist_plot_vary_rho}), which is strongly peaked at $n/N = 1$, indicating that we have a strong condensate. It tells us that typically at an instant in time nearly all of the mass occupies a single site. 

The maintenance of order in the strong condensate phase can be attributed to the dynamics of this tail of mass which trails directly behind the condensate itself. Since a larger hopping rate $b$ biases the rate function $u(n)$  towards hops from sites with a low occupancy $n$, hops from sites with low $n$ will occur much more frequently than from those sites with large $n$. When the condensate hops, it leaves behind a site of occupancy 1. A single unit of mass has the largest possible hop rate, and thus it seems plausible that it is much more likely for the single mass unit immediately behind the condensate to recombine with it than it is for the condensate to hop again and away from the mass it left behind. In this way, it appears that the strong condensation is a consequence of the condensate being unable to escape from the tail of mass trailing behind.  As such, the structure of a very strongly occupied condensate and its very short tail of a few masses is maintained as they move through the system.  We provide further evidence for this intuitive picture of strong condensation within a theoretical treatment below.

\section{Analysis of the moving strong condensate}

The fact that the system exhibits a coherent moving structure, comprising a condensate and its tail, implies that the occupancies of the sites near the condensate are likely to be correlated. An exact solution for the stationary distribution of this model is therefore unlikely to be easily attainable. In order to construct an approximate theory that allows us to estimate the critical value $b_c$ at which the transition to a strong condensate occurs, we make two main assumptions.  First, we work in the frame of reference of the condensate, labelling sites $k=1,2,3,\ldots$ according to how far \emph{behind} the condensate they are, and assuming that almost all of the mass occupies site $k=0$ (i.e., that a strong condensate has formed).  Second, we allow the probability distribution for the number of particles $n$ on each site $k$ to take a \emph{different} form, $p_k(n)$, on each site \emph{but}, to allow analytical progress, we assume that the occupancies on different sites are uncorrelated.

We distinguish between the dynamics of particles in the tail, and of the condensate itself.  When mass is transferred in the tail (either by a single particle hop, or by a backchip), it moves in the \emph{negative} $k$ direction, \emph{towards the condensate}, from $k+1 \to k$.  This leads to a mass current $J_k$ due to hopping from site $k$ to $k-1$ given by
\begin{equation}
\label{eq:Jk}
  J_k = u(1)p_k(1) + \sum_{n=1}^\infty (n-1)u(n)p_k(n) \;.
\end{equation}
Meanwhile, the condensate hops with rate $\simeq 1$ since its mass is of order $N$ and we consider the limit of large $N$. In the frame of reference of the condensate, this causes the whole tail to shift its position which is accounted for by relabelling the indices $k \to k-1$.  Consequently, the \emph{total} average current \emph{arriving at site $k$} in the \emph{positive $k$ direction} {\it i.e.} from site $k-1$ to $k$ is
\begin{equation}
\label{eq:Kk}
  K_{k-1} =  \overline{n_{k-1}} - J_{k} \;.
\end{equation}
By continuity, the mean occupancy of site $k$ changes with time as
\begin{equation}
  \frac{\D}{\D t} \overline{n_k} = K_{k-1} - K_k \;.
\end{equation}
In the steady state $\frac{\D}{\D t} \overline{n_k} = 0$, so we find $K_k = K$ for all sites $k$. Thus, \eref{eq:Kk} becomes
\begin{equation}\label{eq:K_nJk}
  K =  \overline{ n_{k-1} }  - J_{k}  \;.
\end{equation}

\begin{figure}
  \centering	
  \includegraphics[width=0.6\linewidth]{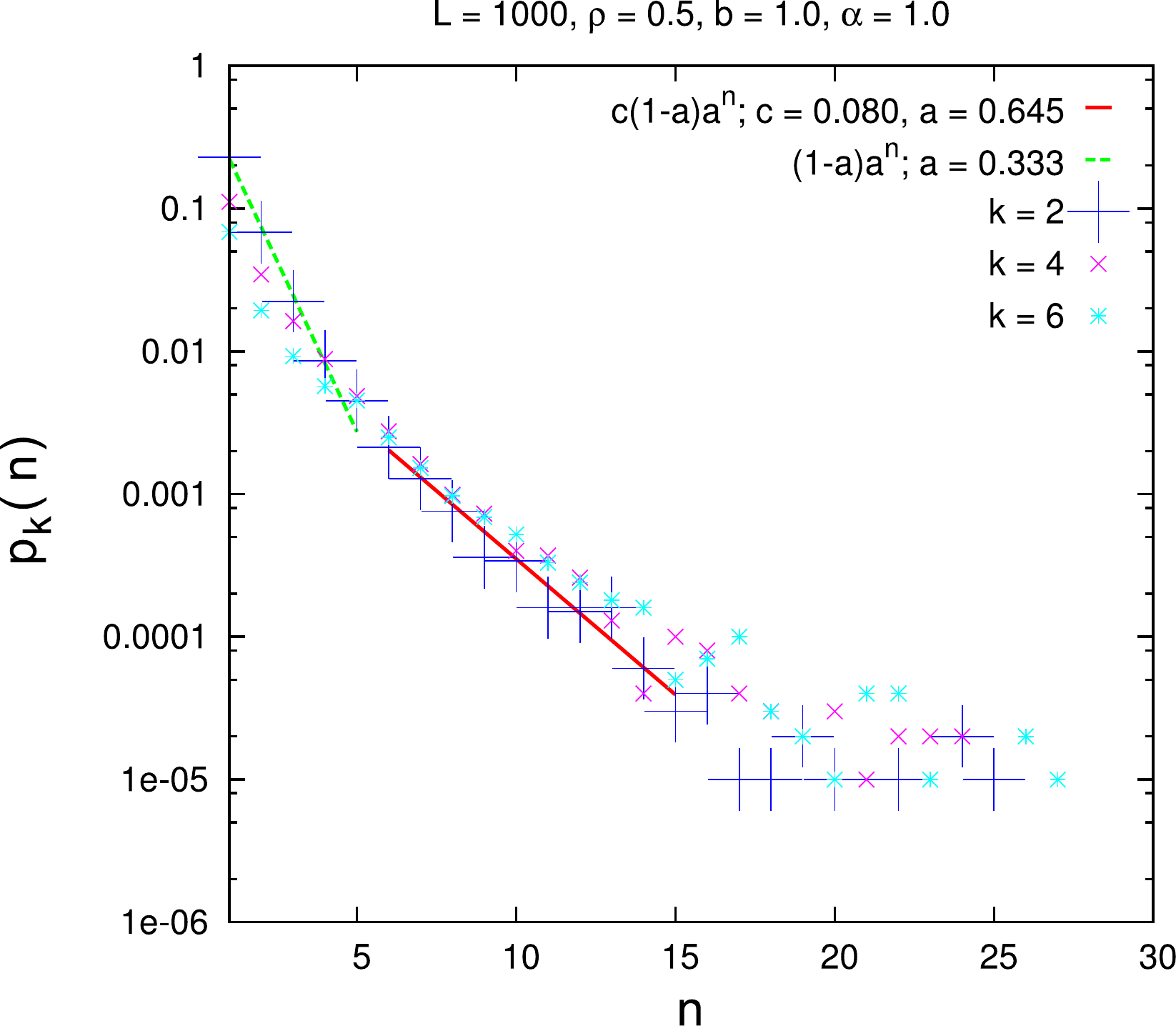}
  \caption{Plots of the probability distributions $p_k(n)$ for having $n$ particles at site $k$ behind the condensate. Here we have plotted the data for the sites closest behind the condensate, and to the measured distribution for $k = 2$ (large blue $+$) have fitted the function $c(1-a)a^n$ (red solid line) to the middle of the tail, and $(1-a)a^n$ (green dashed line) to the front of the tail. (Similar fits can be made for other values of $k$.) Above $n\sim15$ the data is too noisy to be fitted to reliably. The data shown is for sites $k = 2,4,6$ and a system with $\rho = 0.5$, $L = 1000$, $b = 1.0$, $\alpha = 1.0$.} 
  \label{fig:k_tail_fits}
\end{figure}

Inserting the explicit form \eref{eq:hop_rate} for $u(n)$ (with $\alpha =1$) into \eref{eq:Jk} we obtain
\begin{equation}
J_k = (1+b)p_k(1) + \overline{n_k} -(1-b)(1-p_k(0)) -b \widehat{n^{-1}_k}
\label{J2}
\end{equation}
where
\begin{equation}
\widehat{n^{-1}_k} = \sum_{n=1}^{\infty} \frac{p_k(n)}{n} \;,
\end{equation}
that is, an average of $1/n$ over the part of the distribution where $n>0$. To proceed we must also determine an appropriate form  for $p_k(n)$. We have performed some numerical analysis of $p_k(n)$ in the sites immediately preceding the condensate to allow us to make the appropriate choice. As shown in \fref{fig:k_tail_fits}, we find that it is not easy to fit a simple function to the distribution $p_k(n)$, but to make progress we assume 
\begin{equation}
  p_k(n) = (1-a_k)a_k^n \;,
  \label{gd}
\end{equation}
a geometric distribution, which describes the mass in different parts of the tail of the condensate reasonably well (\fref{fig:k_tail_fits}). This is much easier to work with analytically than other possible assumed distributions as it has the useful property that the parameter $a_k$ can be expressed in terms of the mean occupancy $\overline{n_k}$ at site $k$ as $a_k = \overline{n_k} / ( 1 + \overline{n_k} )$. This allows us to express the current \eref{J2} entirely in terms of $\overline{n_k}$ and $b$, using
\begin{eqnarray}
p_k(0) &=& \frac{1}{1+ \overline{n_k}} \\
p_k(1) &=& \frac{\overline{n_k}}{(1+ \overline{n_k})^2} \\
\widehat{ n_k^{-1} } &=& \frac{\ln(1+ \overline{n_k})}{1+ \overline{n_k}} \;.
\end{eqnarray}
Inserting these expressions into \eref{J2}, we find
\begin{equation}
  J_k = \overline n_k   + \frac{(1+b)\overline n_k}{( \overline n_k + 1)^2} 
- \frac{(1-b)\overline n_k}{( \overline n_k + 1)} 
- \frac{b \ln(\overline n_k + 1)}{(\overline n_k + 1)} \;.
\end{equation}

Next, we make the continuum approximation $k\to x$ in \eref{eq:K_nJk} by Taylor expanding $\overline{n_{k-1}}$ about $x=k$ to first order. This leads to the equation
\begin{equation}
\frac{\partial \overline{n}}{\partial x} = f(\overline{n}) - K 
\label{flow}\end{equation}
where
\begin{equation}\label{eq:fn}
  f(\overline{n}) = 
  - \frac{(1+b)\overline n}{( \overline n + 1)^2} 
  + \frac{(1-b)\overline n}{( \overline n + 1)} 
  + \frac{b \ln(\overline n + 1)}{(\overline n + 1)} \;.
  \label{f}
\end{equation}
The boundary condition is $n_0 = 1$ which comes from the fact that every time the condensate hops it leaves one particle behind. In the continuum limit, this boundary condition becomes $\overline{n}(0)=1$.
\begin{figure}[h!]
  \centering
  \subfloat[]{\label{fig:nclessn0} \includegraphics[width=0.4\linewidth]{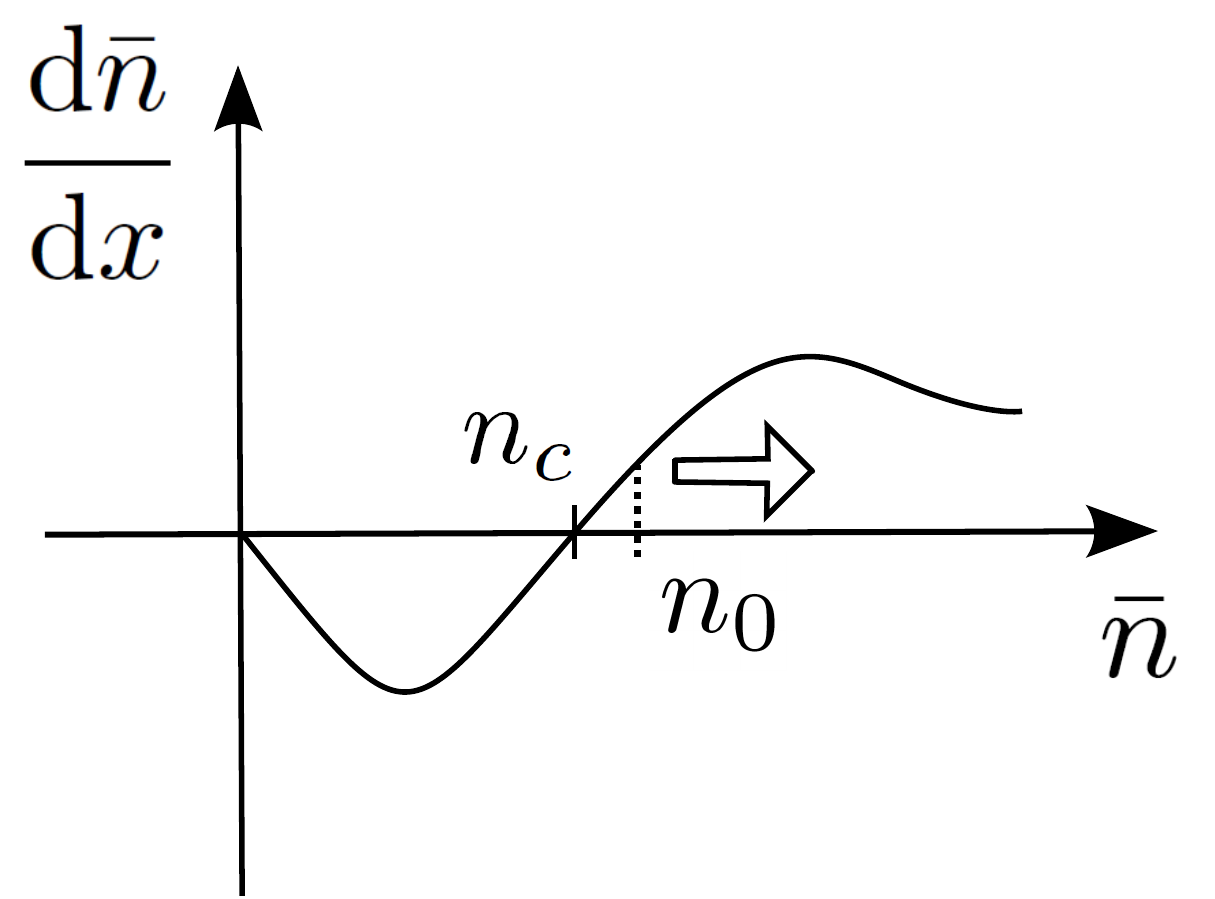}}\hspace{0.1\linewidth}
  \subfloat[]{\label{fig:ncgtrn0} \includegraphics[width=0.4\linewidth]{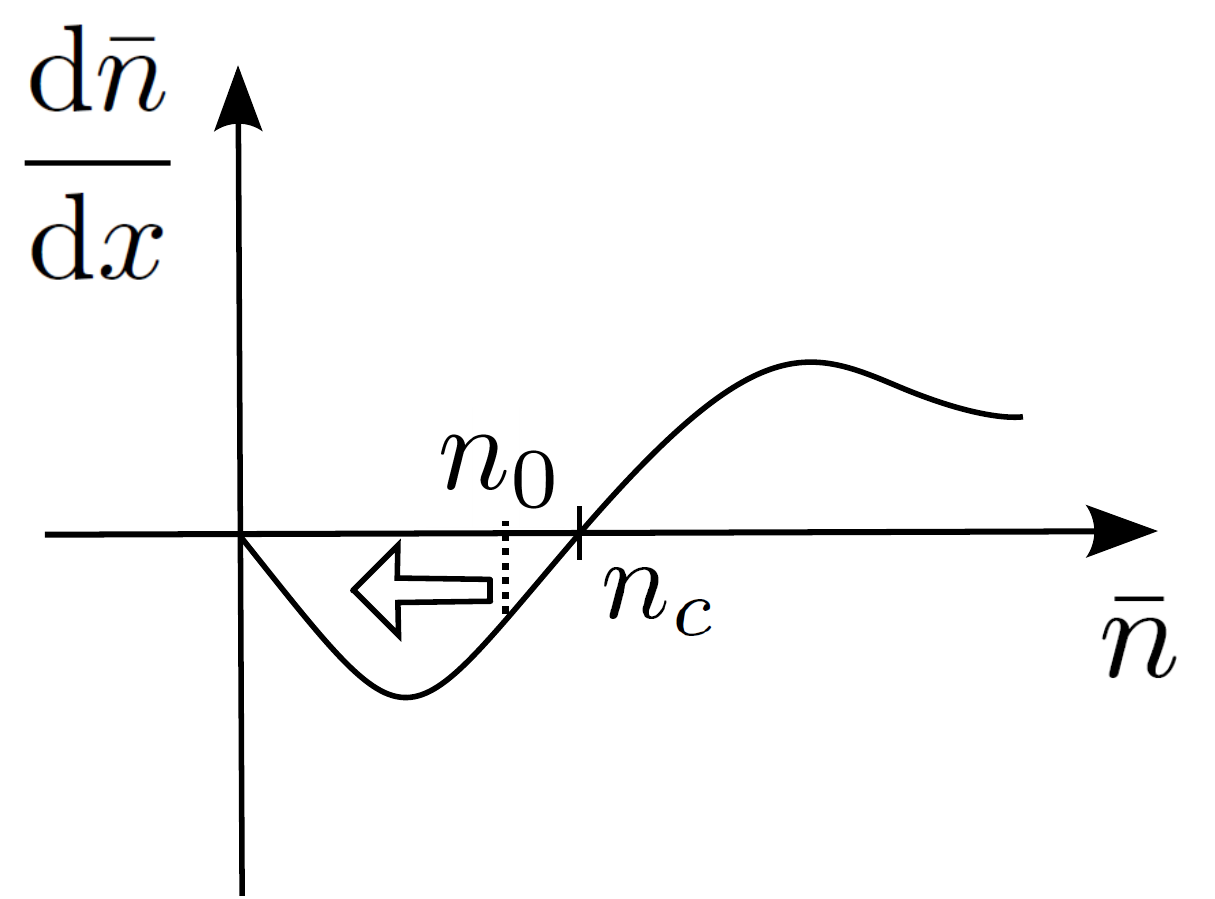} }
  \caption{Using the boundary condition $\bar{n}(x=0) = n_0 = 1$ we see graphically that (a) if $n_c < n_0$ then as $x$ increases, as we move further away from the condensate, so too does $\bar{n}$. It does so indefinitely, resulting in an infinite mean occupancy infinitely far from the condensate. (b) If $n_c > n_0$, then we see that as $x$ increases, $\bar{n}(x)$ decreases to $0$. }
  \label{fig:dndx_sc}
\end{figure}

For the case of the strong condensate, there is no mass at $x\to\infty$ which means that   $K=0$  and therefore that $\frac{\partial \overline{n}}{\partial x} = f(\overline{n})$. 
The form of $f(\overline{n})$ is illustrated in \fref{fig:dndx_sc}.
We note the limits $f(0)=0$ and 
$f(\overline{n})\to 1-b$ for $\overline n \to \infty$. 
We also observe that there is an unstable fixed point at $n_c$, the non-zero root of $f(\bar{n})$,  at which $\frac{ \D \bar{n}}{\D x} = 0$. As illustrated in \fref{fig:dndx_sc}, the value of $n_c$ relative to the boundary condition $n_0 = 1$ will iteratively determine the values of $\bar{n}(x)$ at successively larger $x$. 

If $n_c < n_0$ (\fref{fig:nclessn0}) then  $\frac{\D \bar{n}}{\D x} > 0 $ for all $\bar{n} > n_c$.  This means that $\bar{n}$ will increase indefinitely as $x$ increases, resulting in an infinite mean occupancy far from the condensate. This  contradicts our assumption that there is no mass as $x\to\infty$ and therefore we discard this solution as unphysical. 

If $n_c > n_0$ (\fref{fig:ncgtrn0}) then the gradient of $\bar{n}$ is negative, and it remains negative up to the stable fixed point at $\bar{n} = 0$. This means that successively further from the condensate $\bar{n}$ decreases continuously to $0$. This is the physical solution. The consistency condition for this solution gives us a condition for the existence of the strong condensate: $n_c > n_0 = 1$. This can be translated into a condition on $b$ by using the fact that $f(n_c) = 0$. Substituting $\overline n= n_0 = 1$ in \eref{f} and setting the resulting expression to zero, we find an equation for the critical value $b=b_c$ such that we have a strong condensate for $b> b_c$. The result is
\begin{equation}
 b_c = \frac{1}{3 -2\ln2} \simeq 0.62 \;.
\label{bc}
\end{equation}

This prediction for $b_c$ agrees fairly well with our numerical results displayed in \fref{fig:crossover} and \fref{fig:collapse}. In that figure the sample variance $\sigma^2 = \sum_i\overline{ (n_i/N- \rho)^2}$ of the  occupancy per particle $n/N$ is plotted against $b$ and shown to increase sharply from $\sigma \simeq 0$ to $\sigma\simeq 1$ at a value of $b \simeq 0.5$.  This corresponds to the transition from the fluid, in which the mass is evenly distributed and the sample  variance is small, to  the strong condensate phase in which the sample variance approaches 1. This transition point appears to be independent of $\rho$ and sharpens as the system size increases. 

\section{Classification of the phase transition}

\begin{figure}[h!]
  \centering
  \includegraphics[width=0.6\linewidth]{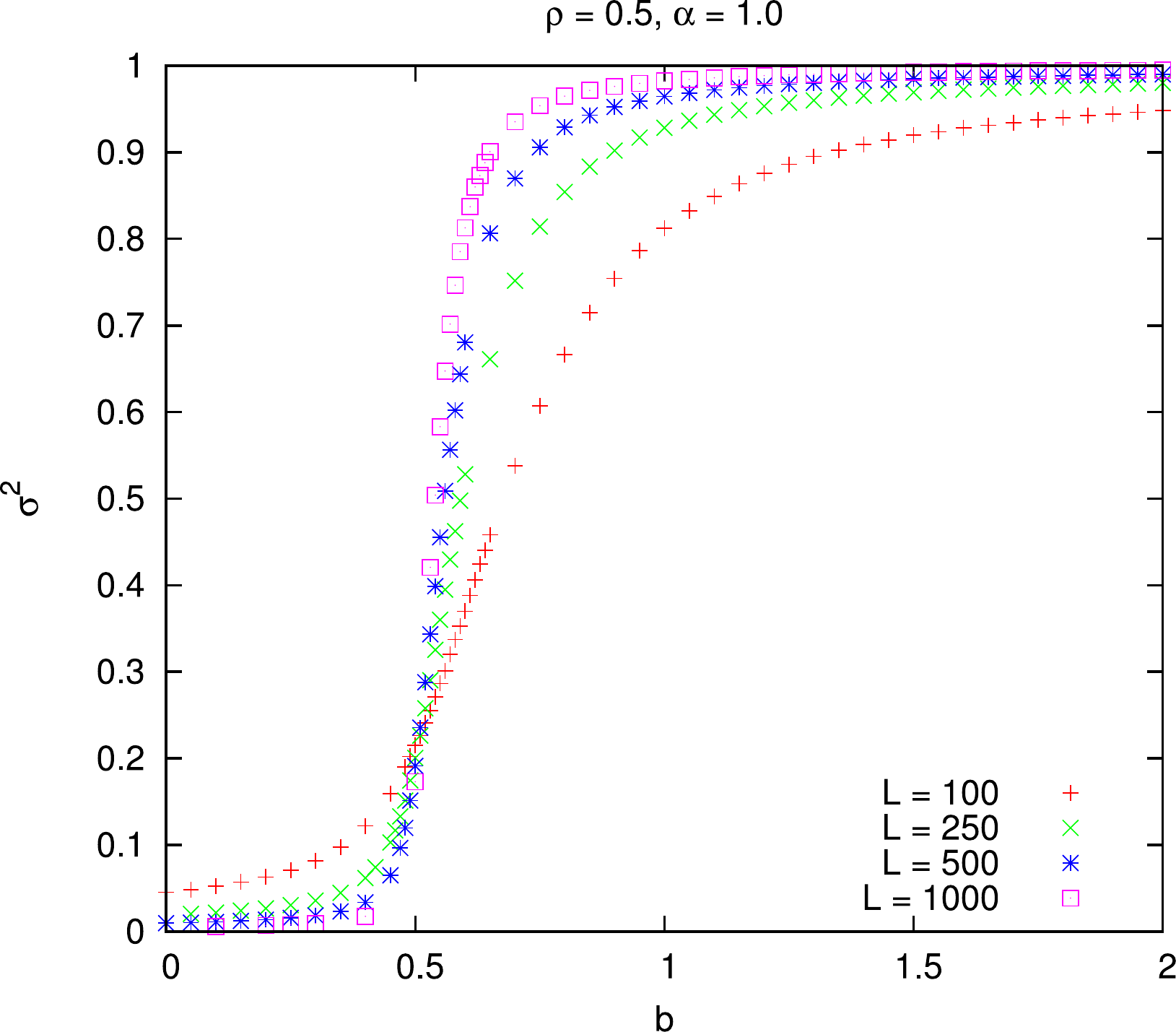}
  \caption{Plots of the order parameter $\sigma$ against the rate parameter $b$, for various system sizes $L$ with density $\rho = 0.5$. A transition is seen to occur for all system sizes, with a clear crossover point in the $\sigma$-$b$ curves at $b = 0.5$, which is indicative of the critical value $b_c = 0.5$. }
  \label{fig:crossover}
\end{figure}
\begin{figure}[h!]
  \centering
  \includegraphics[width=0.6\linewidth]{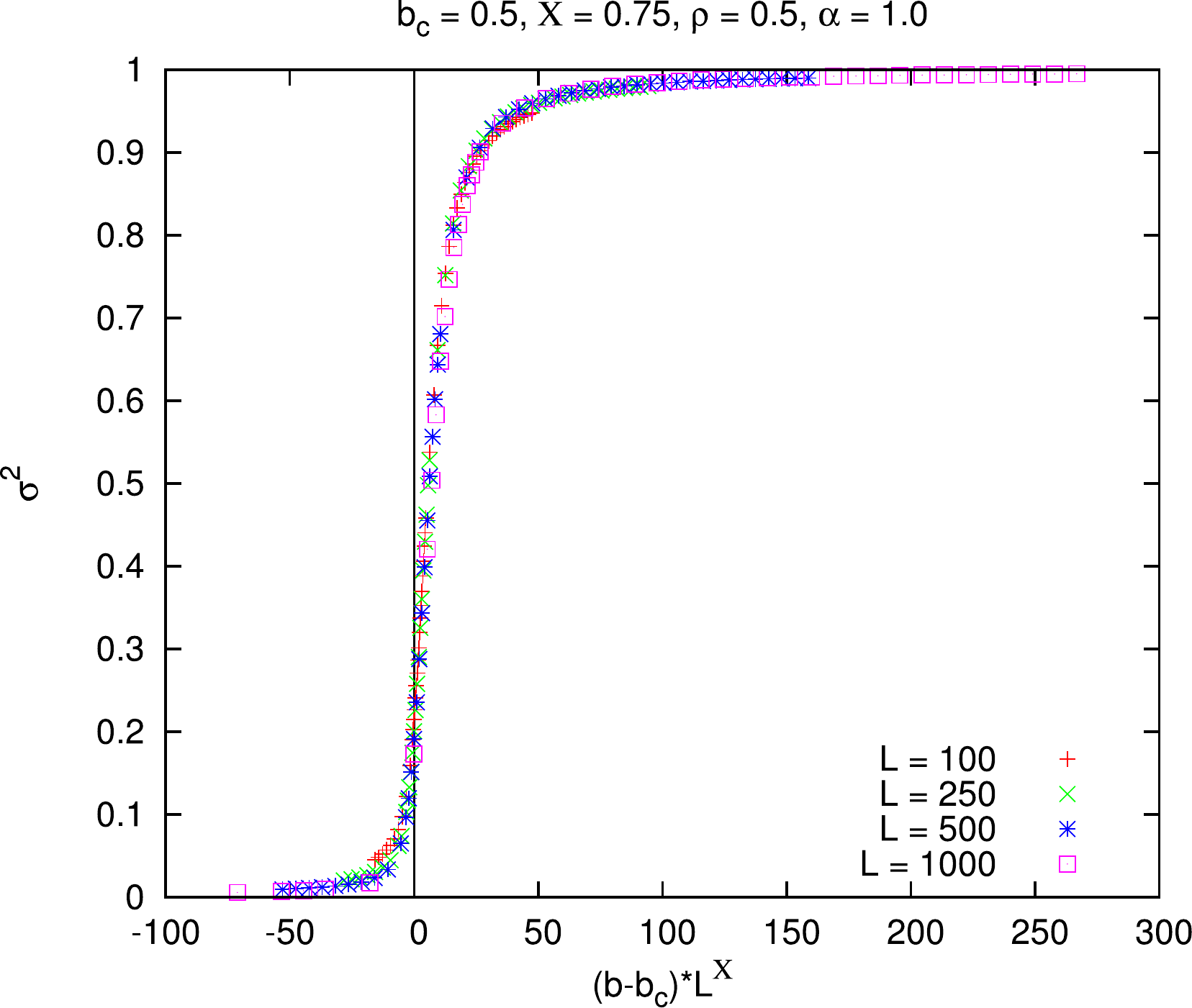} 
  \caption{Plots of the order parameter $\sigma$ against the rate parameter $b$, for various system sizes $L$ with density $\rho = 0.5$. (b) We perform a finite size scaling procedure on $b$, rescaling it to $(b-b_c)L^X$. The choice of parameters for the best collapse of the data onto a single curve is $b_c = 0.5$ and $X = 0.75$.}
  \label{fig:collapse}
\end{figure}

To learn more about the critical value $b_c$ and the nature of the transition we analyse data from different system sizes $L$ at the same density $\rho = 0.5$. First, by studying a plot of the order parameter $\sigma$ against $b$ for this data in \fref{fig:crossover}, we see the transition occurs over a similar range of $b$ for all system sizes $L$. Furthermore, the transition sharpens with increasing system size and there is a stable intersection of the curves at $b = 0.5$. This strongly suggests that in the thermodynamic limit the transition would be discontinuous in $\sigma$ at $b_c = 0.5$. Our confidence in our measurement of $b_c$ is reinforced by the result of applying a finite size scaling procedure to the same data, as shown in \fref{fig:collapse}. We plot the order parameter $\sigma$ against a rescaled hop rate parameter $b^{'} = L^X(b-b_c)$ and, through our choice of $b_c$ and $X$, find the best data collapse when $b_c = 0.5$, and the scaling exponent $X = 0.75$.

\begin{figure}[h!]
  \centering
  \includegraphics[width=0.6\textwidth]{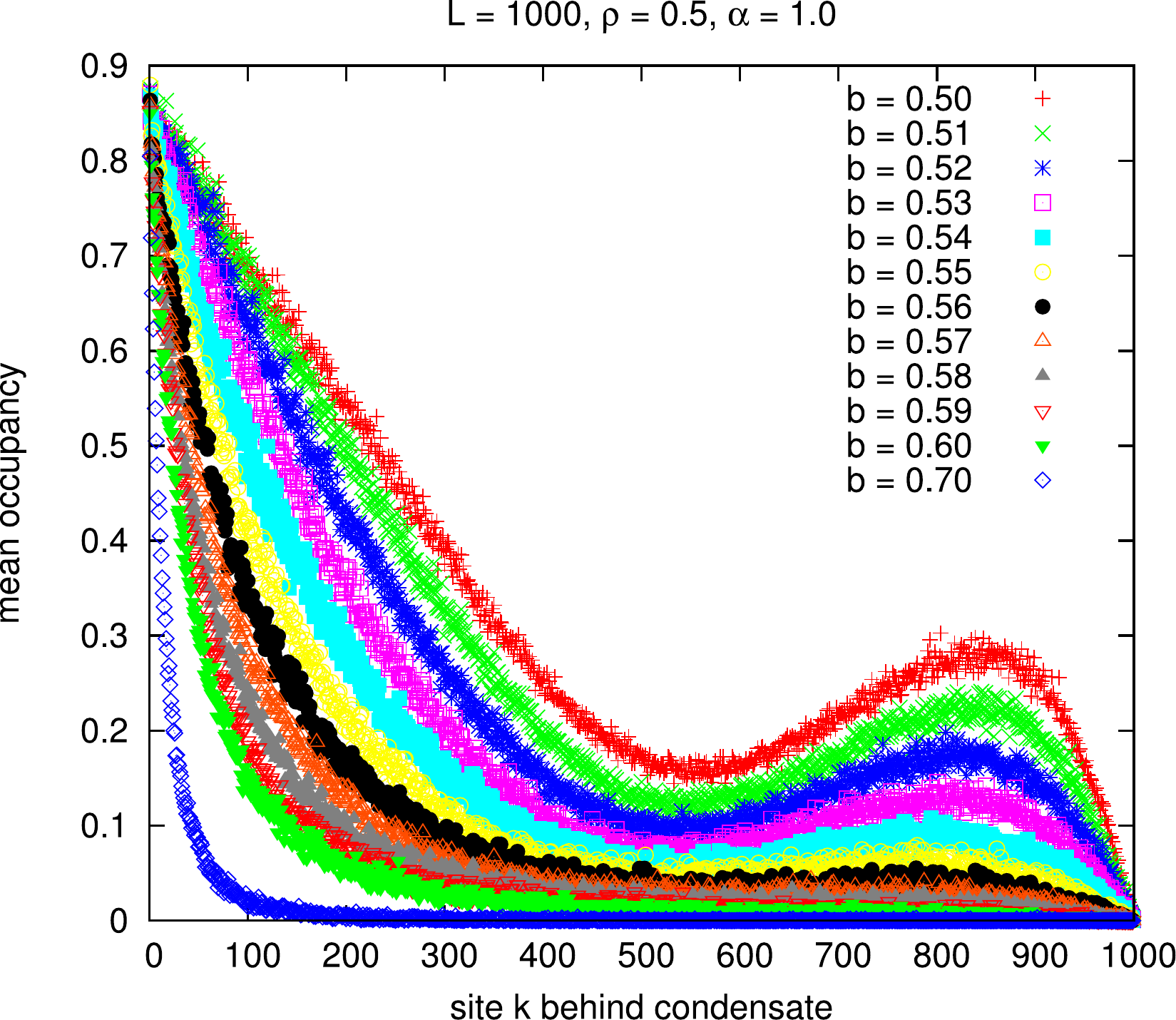}
  \caption{A plot of the mean occupancy at a site $k$ behind the condensate site, measured numerically. The decay length in the region $0<k<500$ can be seen to slowly increase as $b\searrow 0.5$.}
  \label{fig:tailshape}
\end{figure}
Interestingly, looking at the distribution of mass at a site $k$ behind the condensate (\fref{fig:tailshape}) we see that there is a decay length associated with the average shape of the tail of mass behind, which increases as $b \searrow 0.5$. To better understand how this decay length changes as the $b$ approaches the (numerical) critical value $b_c$ from above, we fit an exponential distribution to the tail at different values of $b$, and then plot the dependence of the fitted decay length $\lambda$ on $b$.

\begin{figure}
  \centering
  \includegraphics[width=0.6\linewidth]{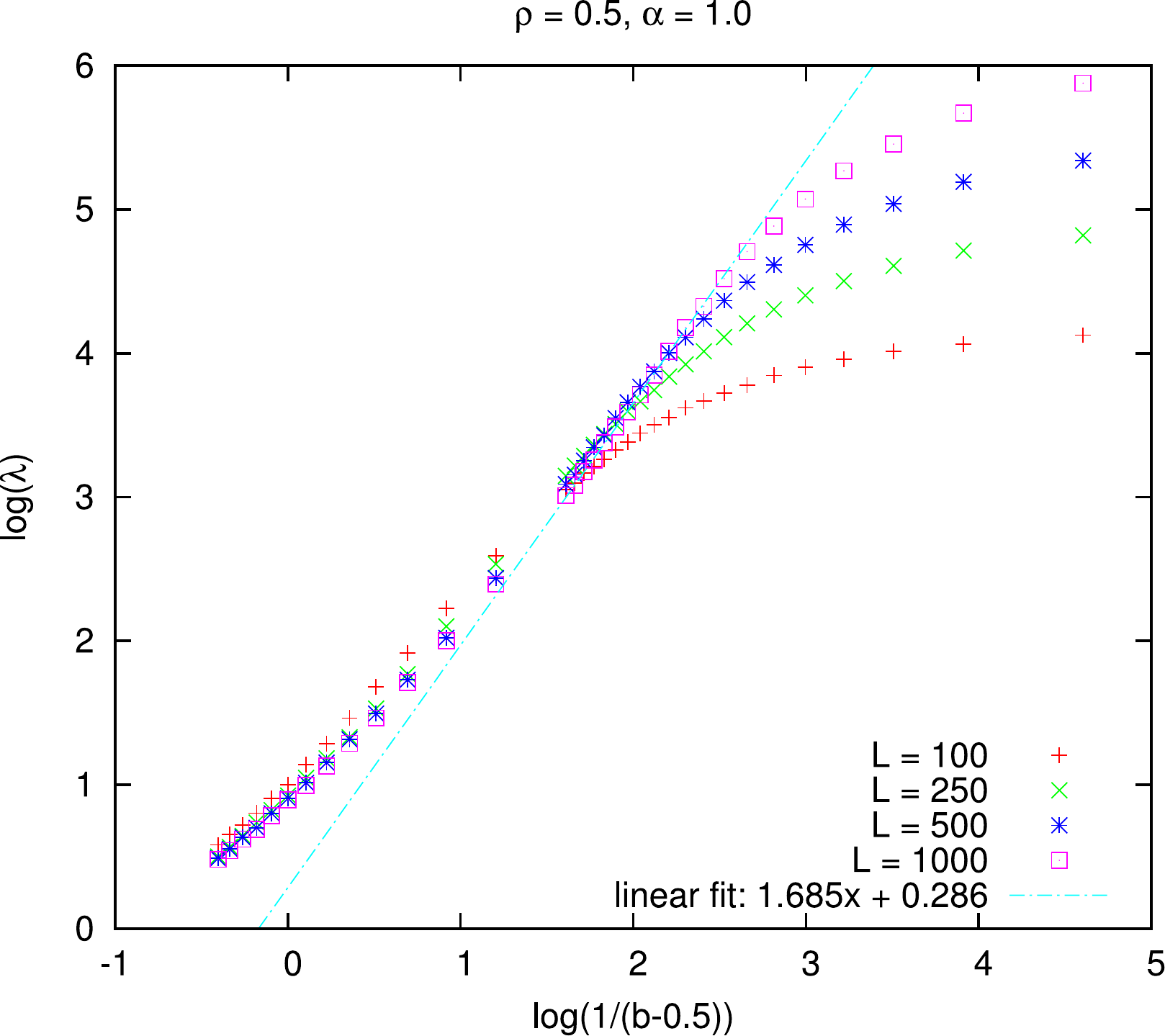}
  \caption{The decay length $\lambda(b)$ measured from the tails in \fref{fig:tailshape} fits a power law in the scaling regime but `flattens out' as $b\searrow 0.5$. Numerically, the system size only affects the $b$ dependence of $\lambda$ as $b$ approaches $0.5$. As the system size is increased the $\lambda(b)$ at $b\searrow0.5$ becomes closer to the power law.}
  \label{fig:flatten}
\end{figure}

As shown in \fref{fig:flatten} the decay length $\lambda$ appears to diverge like a power law as $b\searrow0.5$ before beginning to flatten off as $b-0.5$ becomes very small. We attribute this flattening off effect to the finite system size: in \fref{fig:tailshape} it is clear that as $b\searrow0.5$ the tails develop some additional structure far from the condensate (in the region $500 <k< 1000$), and in \fref{fig:flatten} the data points move closer to the power law fit as the system size is increased in such a way that we expect the infinite system would show the power law divergence without the flattening out effect. 

A diverging length scale of this nature is a characteristic of second order phase transitions, seemingly at odds with the early evidence for the transition being first order in nature. The resolution is to conclude that the phase transition here exhibits the characteristics of a \emph{mixed order} or \emph{hybrid} phase transition. Transitions of this nature are not unprecedented,
for example, in \cite{Bar2014,Bar2014a} mixed order  transitions in long range lattice models are considered.
In another example  a phase transition in the size of the giant viable cluster of a  multiplex network, is shown to be
discontinuous in the order parameter but also to exhibit  critical behaviour above the critical point \cite{Baxter2012}. This asymmetry is attributed to the specifics of the dynamics, which only provide a mechanism for critical behaviour above the critical point and not below.

We speculate that the mixed order transition in the present work may also be attributed to the difference in mechanisms above and below the transition: from above, the transition is brought about by the divergence of a length scale in a coherent structure, namely the tail; from below we see no coherent structures until the condensate itself is formed.

Another interesting observation is that the critical exponent here is measured to be approximately $1.7$, which is similar to the value 1.7338 of the numerically measured critical exponent associated with the temporal correlation length in directed percolation (DP)\cite{Hinrichsen2000a}. Similarities can be seen between the dynamics of the mass in the tail in the frame of reference of the condensate and the dynamics of the driven asymmetric contact process (DACP) which exhibits a phase transition in the DP universality class, specifically with the temporal exponent measured at 1.7(2)\cite{Costa2010}. Here, the backchip process generates new singly-occupied sites which, in the language of the DACP, become ``inactive'' by recombining with other occupied sites, over timescales which are short compared to the timescales at which mass moves from greater-than-singly occupied sites. It may be of interest to investigate such potential similarities in greater depth in the future.

\subsection{Divergence in the approximate theory}

\begin{figure}[h!]
  \centering
  \includegraphics[width=0.6\textwidth]{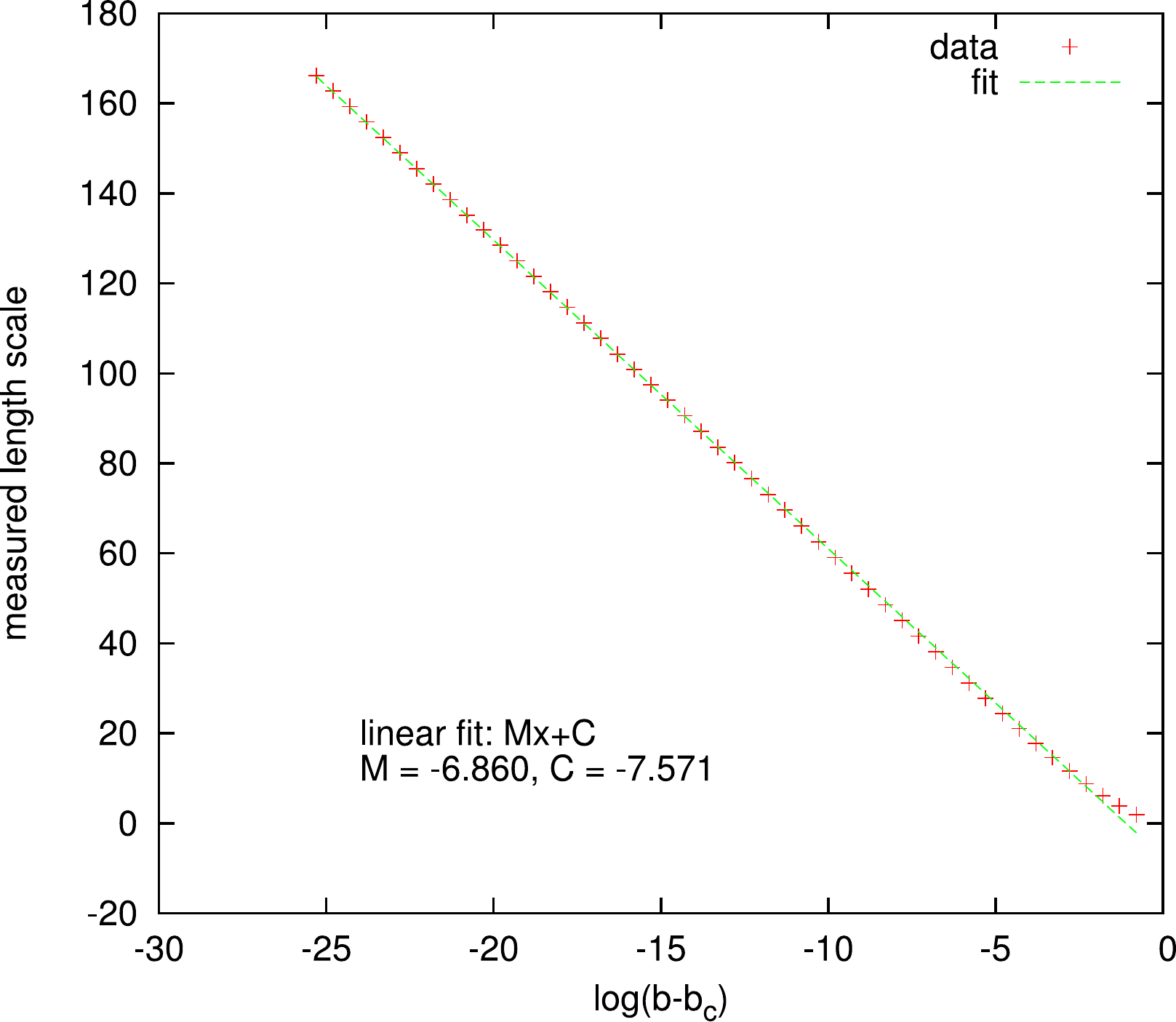}
  \caption{By numerically integrating \eref{f} we measure a length scale in the tail to diverge logarithmically as $b\searrow b_c$. The length scale is defined as the distance from $x=0$ to the position behind the condensate at which the absolute value of $\D n/\D x$ is largest, which was the most distinct and well defined feature of $n(x)$.}
  \label{fig:mf_log}
\end{figure}

To see whether our approximate theory also captures the existence of a mixed order transition we numerically integrated \eref{f} in order to measure how a length scale in the profile of the solution for $n(x)$ changes as $b\searrow b_c$. In contrast to the results from simulation (\fref{fig:flatten}), we find (\fref{fig:mf_log}) that according to the theory the length scale $\lambda(b)$ of the tail diverges as 
\begin{equation} 
  \lambda(b) \sim \log \left( \frac{1}{b-b_c} \right)\;.
\end{equation}
The divergence is still indicative of a mixed-order transition, but it points to one which  has a weakly diverging length scale.

This can also be seen in the approximate theory by studying the mass in the tail very close to the condensate. By considering $\epsilon = b - b_c \ll 1$  and $\eta = n_0 - \bar{n} = 1 - \bar{n} \ll 1$ we can make a Taylor expansion of $f(\bar{n}, b)$ given in \eref{eq:fn}, to find
\begin{equation}
  f(1-\eta, b_c + \epsilon) = f(1,b_c) - A \epsilon + B \eta  + \mathcal{O}(\eta^2, \epsilon^2, \epsilon\eta) \;,
\end{equation}
where
\begin{equation}
  A = \frac{1}{4b_c}\;,\quad B = \frac{1 - b_c\ln2}{4}\;,\quad\mbox{and}\quad f(1,b_c) = 0\;.
\end{equation}
Using the relationship $f(\bar{n},b) = \frac{\D \bar{n}}{\D x}$ we can then write
\begin{equation}
  \frac{\D \eta}{\D x} = A \epsilon + B \eta\;
\end{equation}
and integrate to find
\begin{equation}
  \eta(x) = \frac{A}{B}\epsilon(\mathrm{e}^{Bx}-1)\;.
\end{equation}
A characteristic length scale $\lambda$ can then be defined by the value of   $x$ at wich  $\eta$ reaches some arbitrary finite value, yielding  $\frac{A}{B}\epsilon \mathrm{e}^{B\lambda }=
\mbox{constant}.$ 
Thus
\begin{equation}
  \lambda(\epsilon) =  \frac{|\ln \epsilon |}{B}
+ \mbox{constant}
\end{equation}
So we see that as $\epsilon \searrow \frac{B\eta}{A}$, the characteristic length scale $\lambda$ diverges slowly as a logarithm. 

\section{Condensation transition for the generic hop rate}

\begin{figure}
  \centering
  \includegraphics[width=0.6\linewidth]{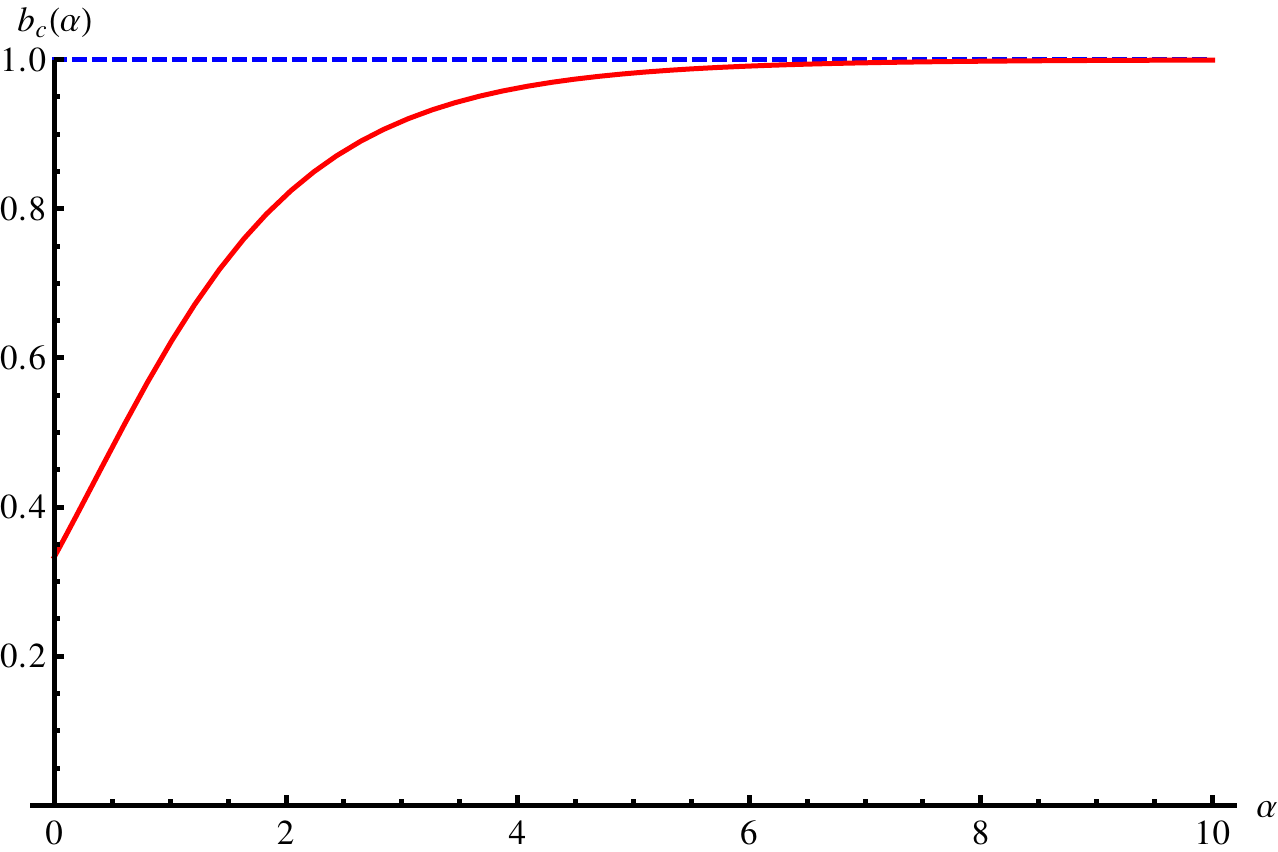}
  \caption{A plot of $b_c(\alpha)$ (red, solid). $b_c$ increases monotonically from $b_c(0)=1/3$ and asymptotically approaches $b_c=1$ (blue, dashed).}
  \label{fig:bc_alpha}
\end{figure}

The transition to a strong condensate phase is found for the generic hop rate \eref{eq:hop_rate} for all $\alpha > 0$. We can repeat the $\alpha=1$ calculation we performed previously but with $\alpha > 0$ and use the definition of the polylogarithm function $ \Li_s(z) = \sum_{n=1}^\infty \frac{z^n}{n^s} $ to find
\begin{eqnarray}
  J_k & = & \overline n_k -1 + \frac{(1+b)\overline n_k}{( \overline n_k + 1)^2} + \frac{1}{( \overline n_k + 1)} \nonumber \\
      & + & \frac{b}{(\overline n_k + 1)} \left[  \Li_{\alpha-1} \left( \frac{\overline n_k}{ \overline n_k + 1} \right) - \Li_{\alpha} \left( \frac{\overline n_k}{ \overline n_k + 1} \right) \right]\;.
\end{eqnarray}
By making the same continuum approximation as before \eref{flow} we find
\begin{eqnarray}
  f(\overline n) & = & 1 - \frac{(2+b)}{( \overline n + 1)} + \frac{(1+b)}{( \overline n + 1)^2}  \nonumber \\ 
      & - & \frac{b}{(\overline n + 1)} \left[  \Li_{\alpha-1} \left( \frac{\overline n}{\overline n + 1} \right) - \Li_{\alpha} \left( \frac{\overline n}{ \overline n + 1} \right) \right]\;.
\end{eqnarray}
Finally, using the  boundary condition $\overline n (0) = 1$ and the constraint on the stable fixed point $n_c(b,\alpha)$, we obtain
\begin{equation}\label{eq:bc_alpha}
  b_c(\alpha) = \left[ 1 + 2 \left( \Li_{\alpha-1} \left( \frac{1}{2} \right) - \Li_{\alpha} \left( \frac{1}{2} \right) \right) \right ]^{-1} \;.
\end{equation}
This function (\fref{fig:bc_alpha}) is monotonically increasing from $b_c(0)=\frac{1}{3}$ and is bounded from above by $1$.  

\begin{figure}[h!]
  \centering
  \subfloat[]{\label{fig:vars_a0-0} \includegraphics[width=0.4\linewidth]{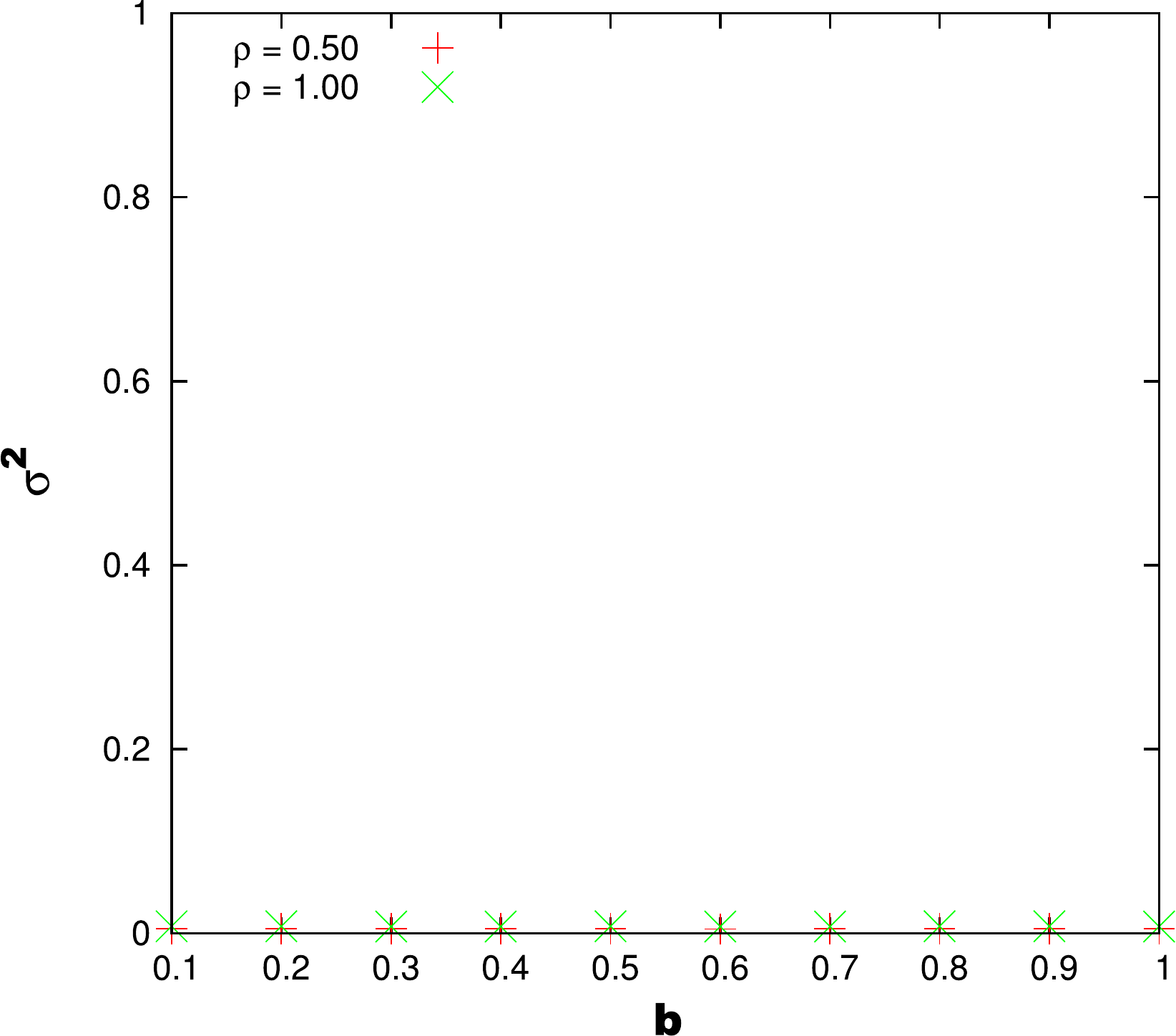}}\hspace{0.1\linewidth}
  \subfloat[]{\label{fig:vars_LOG} \includegraphics[width=0.4\linewidth]{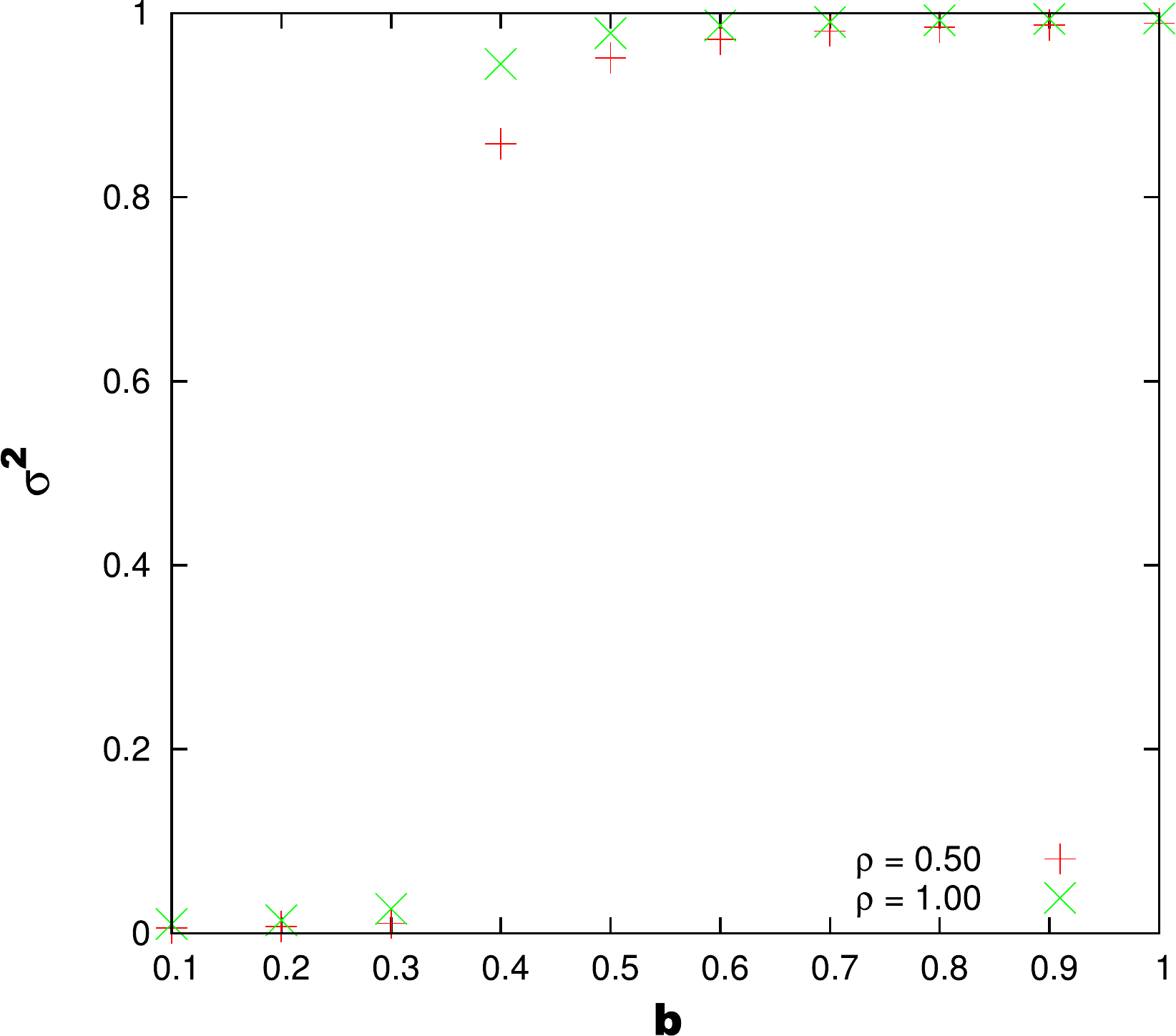} }
  \caption{ Plots of the variance $\sigma^2$ of the occupancy distribution against the rate parameter $b$, for systems of size $L = 1000$. (a) With $\alpha=0$ all masses move with the same rate. Thus there is no mechanism for the formation of the strong condensate, and no transition is observed in $b$. (b) Using the modified hop rate $u_{log}(n)$ given in \eref{eq:u_log} we can probe $b_c(\alpha)$ as $\alpha \searrow 0$. A transition occurs when $b$ is in between $0.3$ and $0.4$, in good agreement with the prediction $b_c(\alpha=0)=1/3$ from \eref{eq:bc_alpha}.}
  \label{fig:vars_0}
\end{figure}

It is important to note that  \eref{eq:bc_alpha} holds for $\alpha>0$ and the point $\alpha =0$ is singular, because for $\alpha>0$ the condensate moves with rate $1$ whereas for  $\alpha = 0$ all masses, including any condensate, move with rate $1+b$. Thus at $\alpha =0$  our mechanism for the maintenance of a moving condensate is no longer valid, as there is no reason small masses would tend to catch up to large masses ahead.  This is confirmed by our simulation results, shown in \fref{fig:vars_a0-0}, for $\alpha = 0$. Measuring the variance of the mass distribution we see no evidence of a condensate forming above a certain value of $b$.

On the other hand, we can probe the validity of \eref{eq:bc_alpha} as $\alpha \searrow 0$ by simulating the dynamics with the modified hop rate 
\begin{equation}\label{eq:u_log}
  u_{log}(n) = 1 + \frac{b}{\ln(n+1)} \;.
\end{equation}
As $\ln n$ increases more slowly than any power of $n$ we can consider \eref{eq:u_log} as approximating the limit of  an arbitrarily small, positive choice of $\alpha$. The results from our simulations with $u_{log}(n)$ presented in \fref{fig:vars_LOG} show that there is a transition to the strong condensate phase in the region $b \sim 0.3-0.4$, which gives us yet more confidence in our analytic result \eref{eq:bc_alpha}. 

\begin{figure}[h!]
  \centering
  \subfloat[]{\label{fig:vars_a2-0} \includegraphics[width=0.29\linewidth]{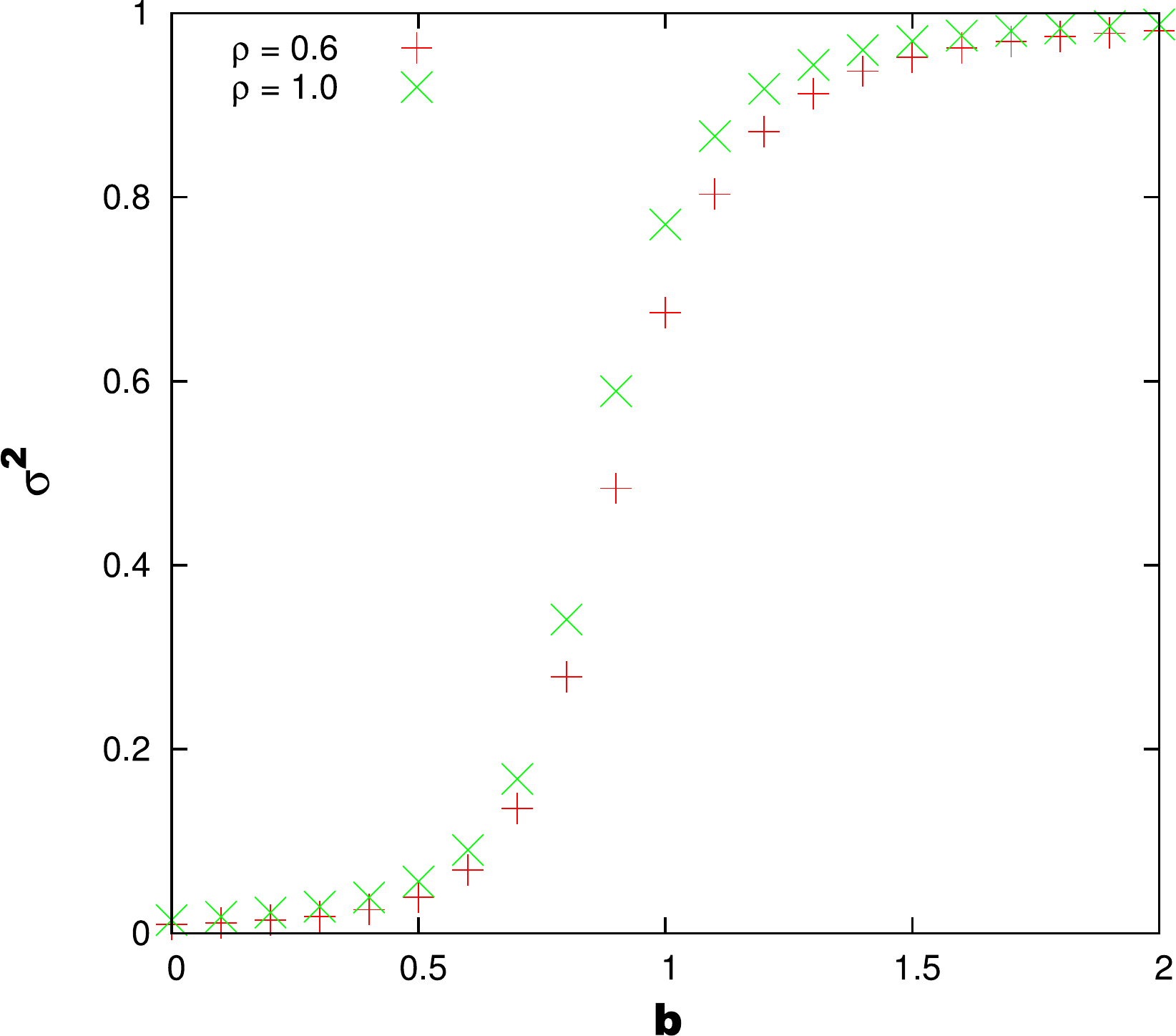}}\hspace{0.04\linewidth}
  \subfloat[]{\label{fig:vars_a10-0} \includegraphics[width=0.29\linewidth]{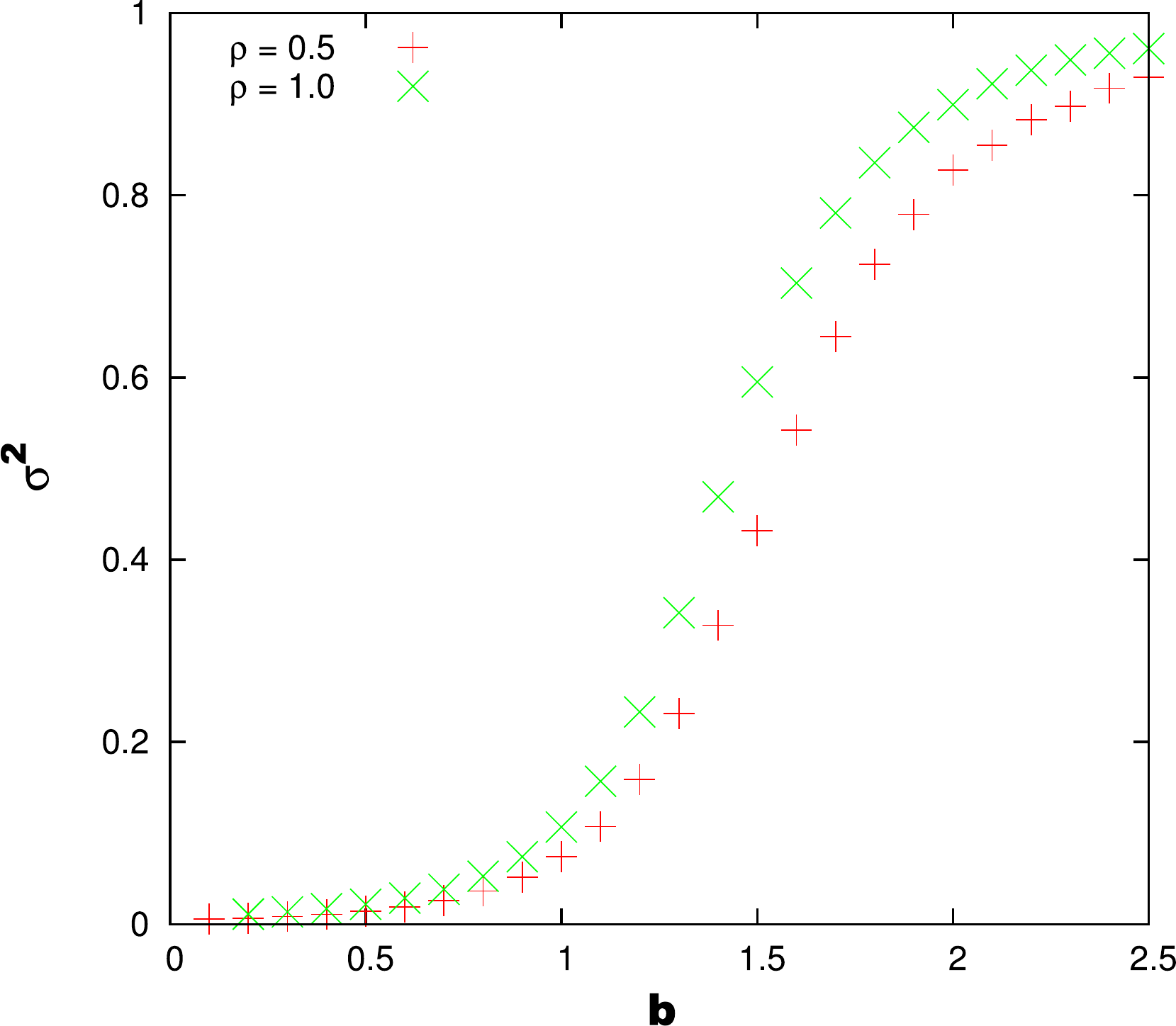}}\hspace{0.04\linewidth}
  \subfloat[]{\label{fig:a10-sys-size} \includegraphics[width=0.29\linewidth]{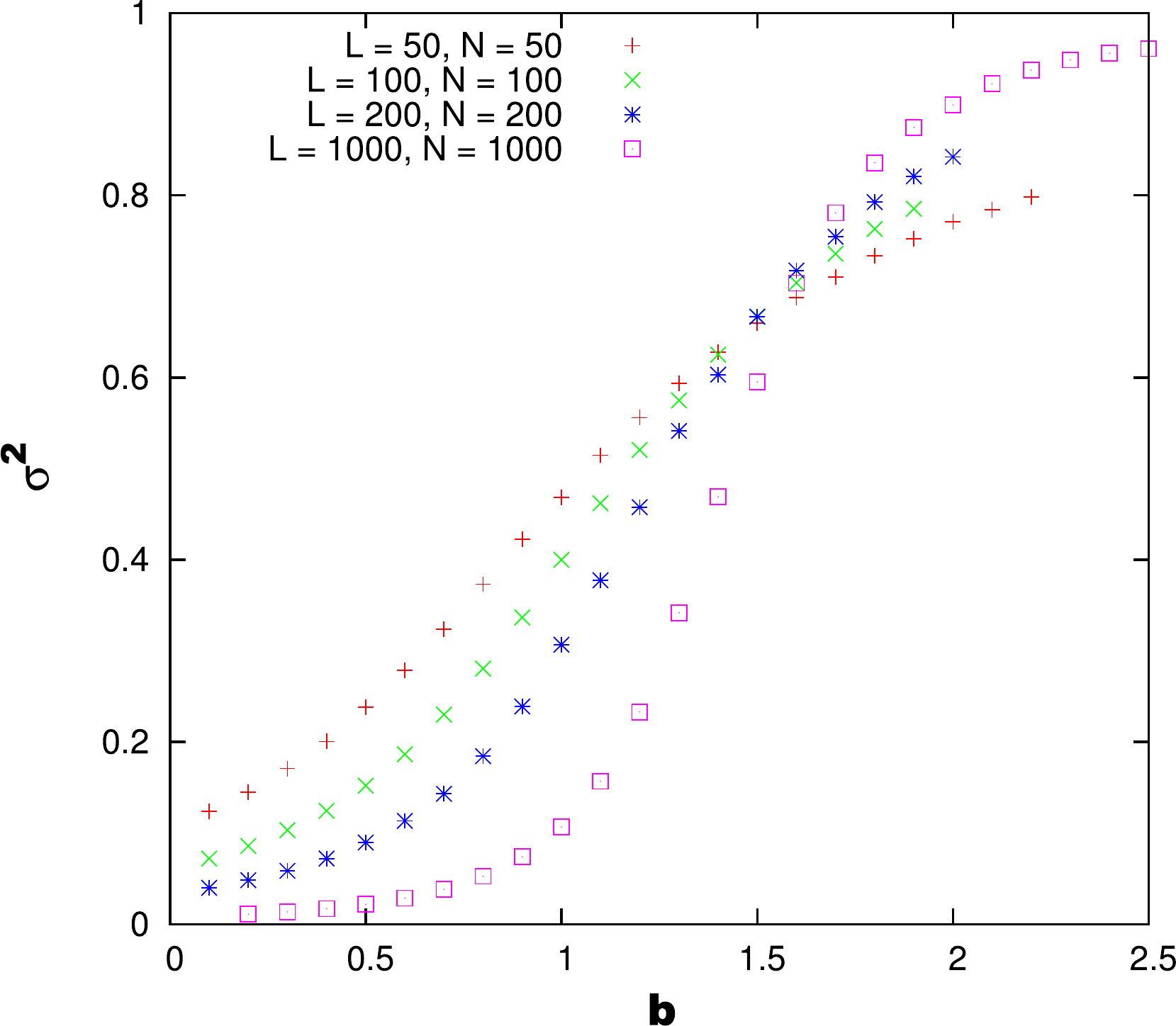}}
  \caption{ Plots of the variance $\sigma^2$ of the occupancy distribution against the rate parameter $b$. (a) $\alpha=2$, $L = 500$. (b) $\alpha=10$, $L=1000$. The formula \eref{eq:bc_alpha} predicts that $b_c(\alpha = 2) = 0.8185$ and $b_c(\alpha=10) = 0.9995$. We find that the transition becomes less sharp as $\alpha$ is increased, and takes place over a range of values of $b$ which are greater than the $b_c$ predicted by \eref{eq:bc_alpha}. (c) $\alpha = 10$. By increasing the system size the transition becomes sharper, which is evidence that its gradual nature is a finite size effect. We can estimate $b_c \sim 1.5 \pm 0.1$ from the crossover of the curves.}
  \label{fig:vars_+}
\end{figure}

We have also performed simulations at $\alpha$ values larger than $1$ (\fref{fig:vars_+}). We find that the transition becomes more gradual for larger values of $\alpha$, and occurs over a region of values of $b$ which are larger than the value of $b_c$ predicted by \eref{eq:bc_alpha}. By studying various system sizes (\fref{fig:a10-sys-size}) we see that evidence that the gradual nature of the transition is a finite size effect, as it becomes more sharp when we increase the system size. The hop rate $u(1) = 1 + b$ for all $\alpha$, but for large $\alpha$ the hop rate from sites with low occupancies (greater than 1) is reduced significantly. This has the effect of suppressing all single occupancy sites because single units of mass always catch up with a site ahead of mass greater than one, in the same way that the condensate is maintained.  We note that although our prediction for $b_c$ seems to agree well with simulations for $\alpha=0$ and $\alpha=1$, for larger $\alpha$, $b_c$ appears to overshoot the asymptote $b_c=1$ predicted by the approximate theory.

\section{Subcritical region}

\begin{figure}
  \centering
  \includegraphics[width=0.6\linewidth]{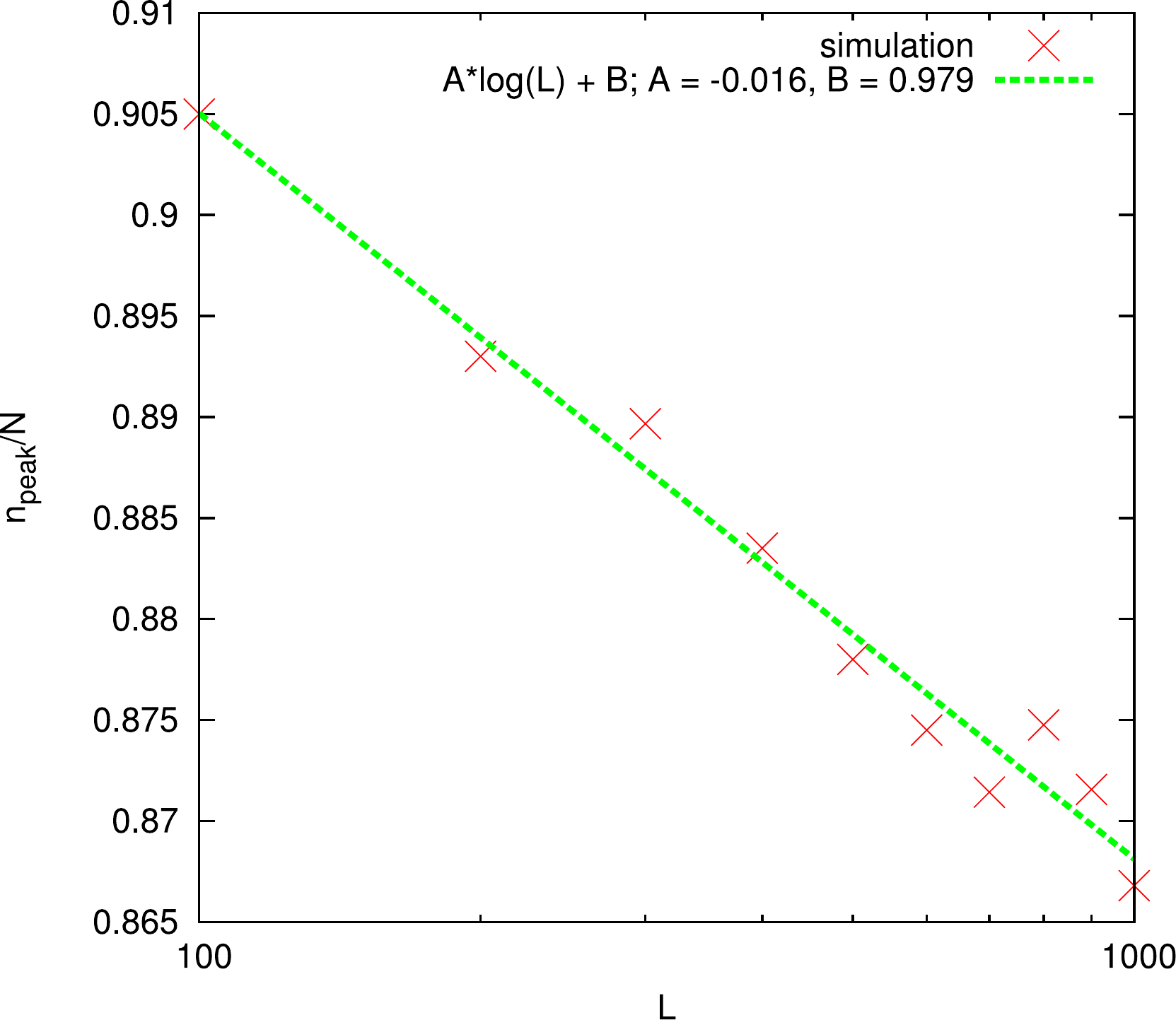}
  \caption{ As the system size is increased, the position of the peak in the probability distribution decreases logarithmically with $L$. This means that the critical density $\rho_c$ has $L$ dependence of the form $\rho_c(L) \sim \ln(L)$. }
  \label{fig:logL}
\end{figure}

Our numerical data and analytical work has shown the existence of a strong condensate phase when $b > b_c$ at all densities. When $b < b_c$ we find a transition from a homogeneous fluid phase when $\rho \lesssim 1$ to a `standard' condensate phase when $\rho \gtrsim 1$. Although this phase quantitatively and qualitatively looks like a standard condensate, numerical analysis shows that the transition density $\rho_c$ diverges as $\ln(L)$, in a similar way to that observed for biased hopping rates in \cite{Rajesh2002}. This can be seen from \fref{fig:logL} where we have analysed the size of the position of the peak, $n_{peak}$, in the distribution of the standard condensate phase. If the fluid contains on average $N_c = L\rho_c(L)$ particles, then $n_{peak}(L) \simeq (\rho-\rho_c(L))L/N = (1 - \rho_c(L)/\rho)$. Our measurements of the value of $n_{peak}(L)$ (\fref{fig:logL}) show that $n_{peak}(L) \propto -\ln(L)$, and thus that $\rho_c \propto \ln(L)$.

\begin{figure}
  \centering
  \includegraphics[width=0.6\linewidth]{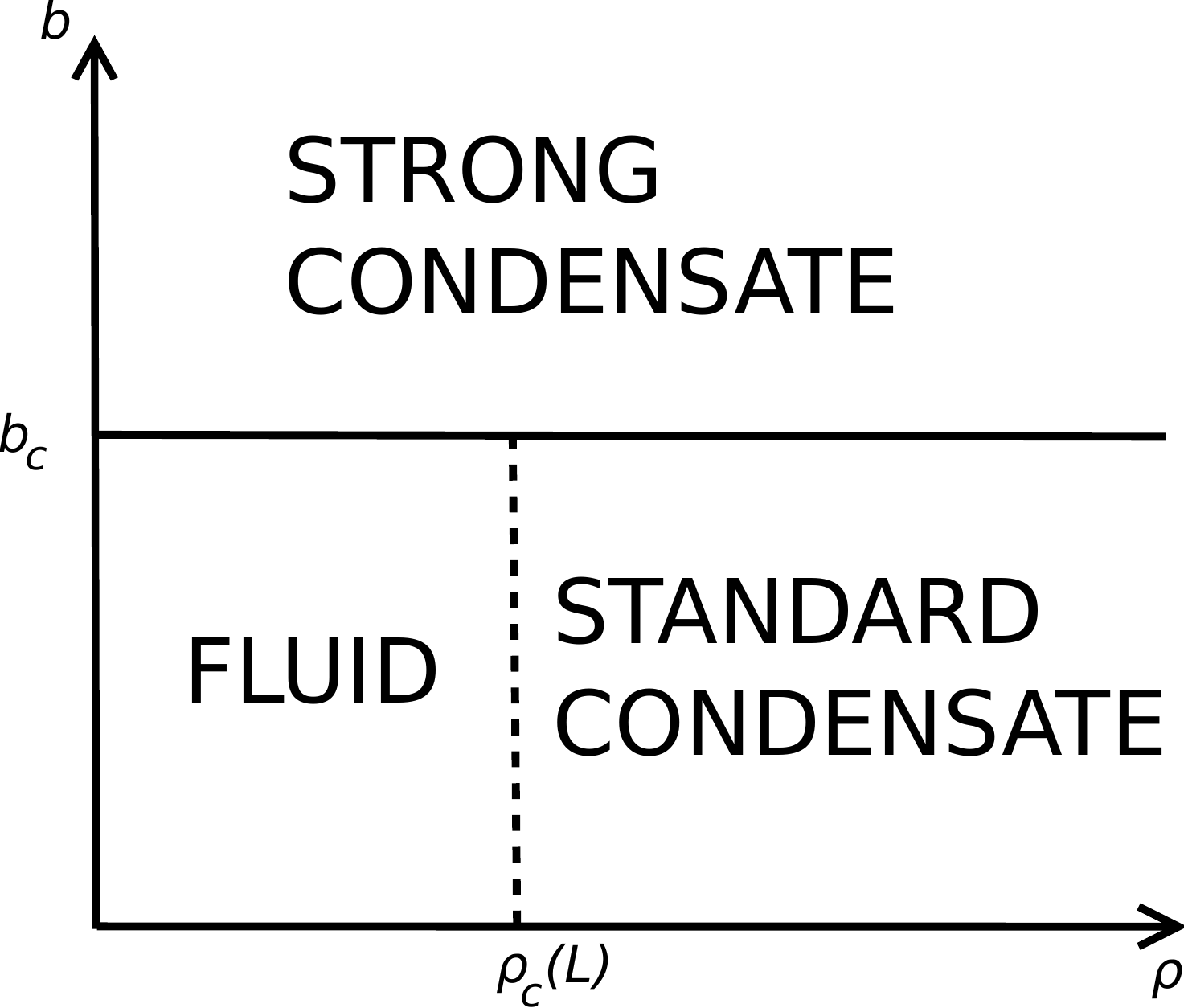}
  \caption{A sketch of the phase diagram for our system. Above the value $b_c$, the system exhibits a strong condensate phase. Below $b_c$ and for low densities the system exists in a fluid phase with a homogeneous distribution of mass throughout. Above a certain value $\rho_c$, the system exhibits the characteristics of a standard condensate phase. Note however that this critical density diverges as $L\to\infty$.}
  \label{fig:phase_diagram_sketch}
\end{figure}

Putting this finding together with the results of the previous sections, we are finally able to sketch a phase diagram for the model in the $\rho-b$ plane. This is shown in \fref{fig:phase_diagram_sketch}.

\section{Conclusion}

In conclusion, the hopping dynamics  invoking `backchip' processes that we have studied in this work give rise to a strong condensate phase in which the condensate and its short tail of trailing particles move together through the system. This phase is present at all densities $\rho$ when the parameter $b$ in the hopping rate $u(n) = 1 + b/n^{\alpha}$ is greater than a critical value $b_c$, which appears to be independent of the system size. Numerically we have measured $b_c = 0.5$ for the case  $\alpha=1$. This is in fairly good agreement with the value $b_c \simeq 0.62$ found using the condensate frame analysis of section 4. We classify the transition as being mixed order as it exhibits a discontinuity in the order parameter $\sigma$, which is indicative of a 1st order phase transition, as well as a diverging length scale, in this case the decay length of the tail of the condensate, which is a characteristic of a 2nd order transition.

Our results also show a number of additional interesting features.  First, the condensate and its tail comprise a coherent object that moves throughout the system, and the stability of the condensate lies in the dynamics of the vanishingly small fraction of particles in the tail.  Once a few particles are left behind through backchipping they quickly rejoin the condensate.
This picture is substantiated by the theory of section 4 which demonstrates that the tail of a moving strong condensate  necessarily decays quickly to zero  for $b$ greater  than a critical value $b_c$.
Second, by extending our analysis to values of $\alpha \ne 1$ in \eref{eq:hop_rate}, we find that the strong condensation phenomenon is generically present for any $\alpha >0$. As illustrated in \fref{fig:bc_alpha} the function $b_c(\alpha)$ increases monotonically from $b_c(0) = 1/3$ and asymptotically approaches $1$ as $\alpha \to \infty$. We recall that in the standard ZRP, condensation is present only for $\alpha<1$. Simulation results confirm the existence of the strong condensate for $\alpha>1$, although the approximate theory appears to underestimate the transition point.

Below the critical $b_c$, we see behaviour more reminiscent of the standard ZRP, in which there is an apparent transition from a fluid phase at low density, to a standard condensate phase above a critical value of $\rho$. However we see numerically that this critical value $\rho_c \sim \ln(L)$ as $L \to \infty$, in a similar way to that observed  in \cite{Rajesh2002}. Here, the condensate is not a true feature of the system in the large $L$ limit,  but rather a finite size effect.  This suggests that systems in which aggregates diffuse and chip, the most relevant quantity in determining whether condensation occurs is the rate of decay of the chip rate with the aggregate size.

To understand the interaction between these dynamical processes better, it would be interesting to study a generalisation of this model where 
$n-a$ particles hop in unison  for $n>a$ and a single particle hops for $n\leq a$.
We have shown here that in the case
$a = 1$ a strong condensate forms and travels through the system, with its structure maintained by the effects of hops from its tail. On the other hand the case $a =N$ yields the zero range process, where only one particle may hop at a time and  a condensation  occurs  for sufficiently large choice of $b$, but with a static condensate.
It would be interesting then to ask
how $a$ should scale with $N$
for one to observe a moving, as opposed to static, condensate and  at what speed it would  travel.

\ack

JW acknowledges financial support from EPSRC (UK) via the SUPA CM DTC.
This work was partially supported by the EPSRC under grant EP/J007404/1.

\section*{References}

\bibliographystyle{iopart-num}
\bibliography{momsc}   

\end{document}